\pdfoutput=1
\documentclass[11pt]{article}
\usepackage[a4paper,margin=1in]{geometry}
\usepackage[T1]{fontenc}
\usepackage{newtxtext,newtxmath}
\usepackage[final]{microtype}
\usepackage{setspace}
\usepackage{amsmath}
\usepackage{graphicx}
\usepackage{booktabs}
\usepackage{tabularx}
\usepackage{longtable}
\usepackage{multirow}
\usepackage{threeparttable}
\usepackage{caption}
\usepackage{enumitem}
\usepackage[section]{placeins}
\usepackage[authoryear,round]{natbib}
\usepackage{xcolor}
\usepackage[colorlinks=true,linkcolor=black,citecolor=black,urlcolor=black]{hyperref}

\setlist{nosep,leftmargin=1.6em}

\title{\textbf{Shared AI Decision Interfaces and Credit Allocation\\
in a Simulated Lending Market}}

\author{
Victoria Ruojie Li\\[0.4em]
{\normalsize Columbia University, New York}\\
{\normalsize \href{mailto:rl3600@columbia.edu}{rl3600@columbia.edu}}
}

\date{}

\begin{document}
\maketitle

\begin{abstract}
\noindent
Language-model agents can choose from a shared menu of actions. I ask whether the order of that menu, which carries no payoff-relevant information, can move credit in a market. Borrowers from three small open-weight model families operate in a simulated collateralized lending market. In each of 12 matched blocks, one market shows all borrowers the same complete action order and the other balances the six orders. An exact randomization test, hash-locked locally before the first model call, rejects the null that common-market net debt does not depend on the assigned order (two-sided p = 2/8,100). The pre-specified lever-up-first versus delever-first estimate is 1.12 units of initial unused borrowing capacity. Every lever-up-first market ended above the balanced market of its block, and every delever-first market ended below it. Exploratory counterfactuals show why the collateral cap does not undo this shock. The cap limits new borrowing, but no rule makes a borrower sell collateral. The market's rules therefore keep 41 percent of the borrowing difference and 93 percent of the repayment difference. Pool-price valuation also returns part of the screened borrowing. Exposure responds more than debt. Lever-up-first markets ended with 58 percent more debt than delever-first markets and about four times the scenario write-offs under a fixed-price shock (exploratory p = 2/8,100). Credit decisions were concentrated as well, because one model family bought 86 percent of the tokens purchased on credit. No market had a liquidation or realized bad debt.
\end{abstract}

\vspace{0.5em}
\noindent\textbf{Keywords:} artificial intelligence; language-model agents; choice architecture; credit allocation; collateral constraints; decentralized finance\\
\noindent\textbf{JEL:} G23, G32, D91, C63, C90

\newpage

\section{Introduction}\label{sec:intro}

A menu lists the same three actions in a different order. No borrower learns anything new. Yet when every language-model borrower in a lending market sees the same order, the market's debt moves toward the action listed first. This paper documents that effect and explains why a collateral constraint does not undo it.

Concerns about AI concentration in finance focus on shared models, data and providers \citep{fsb2024ai,danielsson2022ai,gensler2020deep}. A decision interface is another shared input. The same menu, response schema and parser can serve many agents. The order of a menu carries no payoff-relevant information, so it should not matter for credit. Yet presentation moves credit demand among human borrowers \citep{bertrand2010advertising}. If it also moves machine choices, and those choices reach balance sheets, a shared interface becomes a common shock with no fundamental behind it.

I test this in a simulated collateralized lending market. The borrowers come from three small open-weight model families and decide over 15 rounds. In each of 12 matched blocks, one \emph{common} market shows all 15 borrowers the same complete action order. A paired \emph{balanced} market spreads the six orders across its borrowers. Orders are randomly assigned to common markets. An exact randomization test, hash-locked locally before the first model call, rejects the null that common-market net debt does not depend on the order ($p=2/8{,}100$). The pre-specified contrast between lever-up-first and delever-first markets is $T=1.12$ units of initial unused borrowing capacity $K_0$. Every lever-up-first market ended above its balanced partner, by 0.61 $K_0$ on average. Every delever-first market ended below its partner, by 0.51. Without the balanced baselines, the two groups' common markets differ by 0.828 $K_0$, or 3,767 USDC per market.

Positional bias in LLM answers is already documented \citep{pezeshkpour2024order,zheng2024selectors,dimino2025positional}. The economics lies in what the market does with it. One might expect a collateral cap to absorb a bias in choices. Exploratory fixed-request counterfactuals suggest that the market's rules absorb about half of this one, and unevenly. Debt adjusts on two margins. Lever-up-first markets borrowed more and also sold less collateral to repay. The cap polices only the first margin. Each purchase uses up the buyer's own borrowing room, but no rule makes a borrower sell collateral before liquidation. On the recorded requests, the rules keep 41 percent of the common-market borrowing difference and 93 percent of the repayment difference. The repayment margin therefore carries 24 percent of the common-market gap in requests and 42 percent of the gap in recorded debt.

The way the cap values collateral adds a second effect. A purchase has price impact in the constant-product pool, and the cap values every borrower's collateral at the resulting price. This valuation returns part of the screened borrowing. On the same requests, it raises the common-market debt gap only from 0.688 to 0.828. Yet it creates all 16 positions above 70 percent of reference collateral value in lever-up-first markets. These borrowers comply with the cap and still carry high leverage at an outside price.

The debt gap matters because exposure responds more than debt. No market had a liquidation or realized bad debt, so I evaluate the recorded books under a specified shock. Lever-up-first markets ended with 58 percent more debt than delever-first markets. Their scenario write-offs under a fixed-price shock were about four times as large (630 against 151 USDC per market). An exploratory randomization test rejects order invariance of these write-offs ($p=2/8{,}100$). Write-offs arise only above a threshold, so a higher level of debt pushes more borrowers across it and scenario write-offs rise faster than debt. The order raises the level of leverage. It does not consistently make leverage more unequal across borrowers.

Credit decisions were also concentrated. Three model families were deployed in equal numbers, but Llama bought 86 percent of the tokens purchased on credit and made all collateral sales. Qwen did not trade. The diversity of deployed models therefore overstates the diversity of decisions.

The paper contributes to four literatures. The first studies correlated behavior among AI agents. LLM studies document sensitivity to reordered alternatives \citep{pezeshkpour2024order}, option labels \citep{zheng2024selectors} and positions in investment choices \citep{dimino2025positional}. Trading simulations examine market decisions without collateralized borrowing \citep{lopezlira2025trade}. Work on correlated LLM behavior emphasizes shared instructions, information and capability \citep{hansen2025animal,ross2026better}, and work on algorithmic monoculture studies decision makers who share one algorithm \citep{kleinberg2021monoculture,bommasani2022picking}. I hold books, information, mandates and model families fixed and reassign only the complete action order. The treatment carries no information, applies to a whole market and is measured in executed debt rather than choices. Separate simulations show that news presentation changes which model families submit orders \citep{li2026selective}. My participation result draws a related distinction between nominal diversity and actual decisions.

The second literature links asset prices to collateral capacity \citep{kiyotaki1997credit,geanakoplos2010leverage}. A purchase raises the pool price used in the borrowing constraint, and a sale lowers it. When a constraint depends on prices, one borrower's purchase changes other borrowers' capacity \citep{davila2018pecuniary}. Each of these links is standard, and so is a cap that limits borrowing and not sales. What I add is a measurement of what these rules do to a shock from a shared interface. The order moved both margins at once, and exploratory fixed-request counterfactuals show how unevenly the two pass through. Requests alone therefore misstate both the size and the composition of what reached the ledger.

The third literature studies how order flow shapes financial outcomes on decentralized exchanges. On these venues, arbitrage opportunities arise from shocks to fundamentals and from the price impact of uninformed trades, and they impose costs on liquidity providers \citep{capponi2025liquidity}. Automated market makers leave price discovery to arbitrageurs and liquidity provision to liquidity providers \citep{lehar2025uniswap}. Evidence on priority fees shows that high-fee order flow is more informative about value \citep{capponi2026price}. I hold liquidity provision fixed and change a shared interface input that carries no payoff-relevant information. In my setting the pool price also enters the borrowing constraint, because the cap values collateral at that price.

The fourth literature asks how presentation moves financial choices \citep{madrian2001power,bertrand2010advertising}. Whether a presentation effect reaches debt is a separate question. A repayment nudge studied by \citet{guttmankenney2025semblance} changed choices but not debt, because cardholders often lacked liquid cash. In this market, repayment is funded by selling collateral, so a different choice does reach debt. Constraints and prices then decide how much of it remains.

The paper makes two contributions. First, a shared action order can change executed credit in this simulated market, and the exact randomization test is the confirmatory evidence for it. Second, borrowing and repayment differences do not enter the ledger in equal proportion once requests pass through the market's rules. This second result rests on exploratory fixed-request counterfactuals and accounting splits. The exposure comparisons are also exploratory, and I label each of these where it appears. Section~\ref{sec:design} describes the market and the experiment. Section~\ref{sec:executed} shows that the order moves debt. Section~\ref{sec:transmission} explains why the cap does not undo the shock. Section~\ref{sec:losses} traces the consequences for credit risk, and Section~\ref{sec:implications} draws the implications. The appendices hold the design records, the earlier experiment, the full exploratory analyses and the larger-model studies.

\section{The Lending Market and the Experiment}\label{sec:design}

\begin{figure}[t]
\centering
\includegraphics[width=\textwidth]{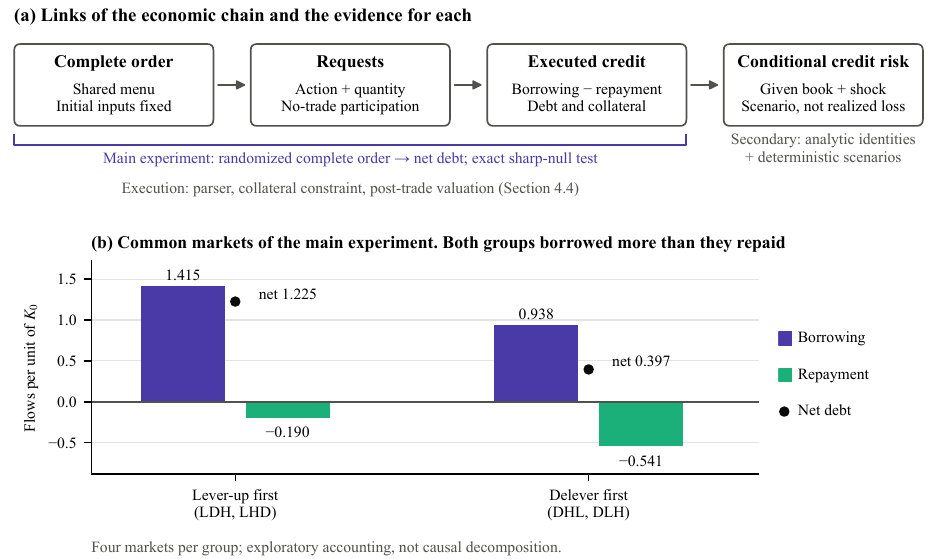}
\caption{\textbf{From action order to executed credit.} (a) The links of the economic chain and the evidence for each. Frozen-state tests measure requests (Section~\ref{sec:requests}). The main experiment randomizes the complete order and tests its effect on net debt (Section~\ref{sec:executed}). Fixed-request counterfactuals trace the recorded requests through execution (Section~\ref{sec:evaluation}). Conditional credit risk is a secondary, deterministic evaluation of recorded books (Section~\ref{sec:losses}). (b) Common-market borrowing, repayment and net debt growth under I1, in units of initial unused reference capacity $K_0$; four markets per order group. Repayment is plotted as negative. The common-market gap is 0.828. The pre-specified $T=1.121$ also includes a balanced-baseline component (Section~\ref{sec:reading}). Means and differences are computed before rounding.}
\label{fig:design}
\end{figure}

A simulation makes this question testable. I can fix initial books, model families, mandates and decision schedules while I assign different complete orders. Borrowers still interact through prices and collateral constraints. I can also follow each request into execution and evaluate the resulting books under common rules. A thin single-collateral pool and a cap valued at the post-trade price make the link between prices and collateral visible. Figure~\ref{fig:design} shows the chain from order to credit and the evidence for each link.

\subsection{The market and its borrowing constraint}\label{sec:market}

Each market has 15 borrowers, a constant-product automated market maker (AMM) with 1,000 SMC collateral tokens and 100,000 USDC, and a lending protocol. The reference price is 100 USDC per SMC. The pool price moves with the price impact of trades, and after each round an arbitrageur closes 25 percent of the log gap to the reference price. Liquidity provision is held fixed. Borrowers start with 5--15 SMC, reference LTVs of 25--55 percent and wallet cash of 900--1,100 USDC. Five agents use each of Llama 3.1 8B, Qwen 2.5 7B and Mistral 7B. Five have each of the opportunistic, neutral and capital-preservation mandates. All act once per round for 15 rounds in a seeded random sequence.

\emph{Lever up} borrows USDC to buy 1--5 SMC and adds the tokens to collateral. \emph{Delever} sells whole collateral tokens and repays debt. Excess proceeds go to the wallet, and a sale at zero debt produces cash rather than repayment. \emph{Hold} leaves the account unchanged. There is no interest, and wallet cash cannot repay debt. When the pool-price LTV exceeds 80 percent, liquidation repays half the debt. It seizes collateral worth the repayment plus an 8 percent bonus and sells it into the pool. The dynamic engine does this mechanically and does not check liquidator profitability or funding.

Let borrower $i$ hold collateral $c_i$ and debt $d_i$, and let the pool hold reserves $(x,y)$. Buying $q$ tokens costs $\beta(q)=yq/[(x-q)(1-\phi)]$, including the AMM fee $\phi=0.3$ percent. The purchase leaves the price at $p'(q)=[y+\beta(q)]/(x-q)$. The engine executes the largest whole quantity $q'\le q$ that satisfies
\begin{equation}\label{eq:cap}
d_i+\beta(q')\;\le\;0.70\,(c_i+q')\,p'(q').
\end{equation}
This \emph{spot-valued cap} constrains new borrowing only. It does not limit collateral sales or existing debt, and no rule requires a borrower to sell collateral before liquidation. These features drive the results in Section~\ref{sec:transmission}. A purchase also raises the pool price and thus the value of every borrower's collateral. Compliance can therefore coexist with a reference LTV $\ell_i=d_i/(100c_i)$ above 70 percent. A reference-valued cap replaces $p'(q')$ with 100, and major lending protocols value collateral with external oracles in this spirit \citep{aave2026v3origin,compound2026docs}. Both market experiments use the same spot-valued engine. The valuation comparisons use fixed-request replay (Section~\ref{sec:evaluation} and Appendix~\ref{app:replay}).

\subsection{The menu and the two interfaces}\label{sec:interfaces}

Prompts show the mandate, the borrower's balance sheet, the pool price and LTV, current news and feedback on the previous decision. The treatment is one of six \emph{complete action orders} of lever-up (L), delever (D) and hold (H). In the revised interface (I1) the menu follows this order. So does the relative order of lever-up and delever in the quantity instruction. Information fields and mandates are the same across orders. Later balance sheets, pool prices and feedback evolve with executed decisions.

The original interface (I0) also ordered the action names in its JSON schema. Its prompt allowed a quantity of zero, but its parser rejected zero for active actions. I1 removes the ordered action list from the schema and makes zero a valid no-trade choice for every action. Invalid responses execute no trade and stay in the denominators. The main experiment uses I1, and an earlier market experiment used I0. Table~\ref{tab:campaigns} and Appendix~\ref{app:expd} document the differences.

\subsection{The outcome}\label{sec:outcome}

The outcome is net debt growth in block $b$ and market $g$.
\begin{equation}\label{eq:y}
Y_{bg}=\frac{D_{bg}^{\mathrm{end}}-D_b^0}{K_{0b}},\qquad
K_{0b}=\sum_i\left(0.70\cdot100\cdot c_{ib}^0-d_{ib}^0\right).
\end{equation}
$K_{0b}$, abbreviated $K_0$, is initial unused borrowing capacity at the reference price. It is about 4,660 USDC per market in the main experiment (3,934--5,411). Paired markets share $D_b^0$ and $K_0$. Net debt growth can exceed this initial capacity because borrowing buys additional collateral. There was no liquidation or write-off in these runs, so net debt growth equals executed borrowing less repayment from collateral sales.

\subsection{Matching markets and assigning orders}\label{sec:randomization}

The main experiment has two six-block stages. Each block pairs a common (C) market, whose agents share one order, with a balanced (H) market that spreads the six orders across agents, two or three per order. Both markets in a block share the same initial books, agent assignments, decision seeds, news and execution schedules. Trades change later agents' prices and states, and that interaction is part of the market outcome.

Within each stage, a uniform assignment $\tau$ gives each of the six complete orders to one of the six common markets. Independently, the balanced allocation $\sigma$ is drawn from 165,888 admissible offset vectors under the inherited slot, family and mandate restrictions. The analysis conditions on $\sigma$. I call a common market lever-up-first, delever-first or hold-first according to the first action in its order. Comparisons across these groups rest on the randomized order. The C and H labels are not randomized, so comparisons between them are descriptive. Appendices~\ref{app:expe} and~\ref{app:history} give the full restrictions and chronology.

\subsection{The test and the status of the evidence}\label{sec:estimand}\label{sec:inference}\label{sec:components}

Let $Y_{bC}(o)$ be a common market's outcome under order $o$, with its inputs and $\sigma$ fixed. The finite-input estimand $\theta$ averages the two lever-up-first orders against the two delever-first orders.
\begin{equation}\label{eq:theta}
\theta=\frac1{12}\sum_{b=1}^{12}\left[\frac12\sum_{o\in\{\mathrm{LDH},\mathrm{LHD}\}}Y_{bC}(o)
-\frac12\sum_{o\in\{\mathrm{DLH},\mathrm{DHL}\}}Y_{bC}(o)\right].
\end{equation}
The pre-specified statistic uses the paired differences $\delta_b=Y_{bC}-Y_{bH}$.
\begin{equation}\label{eq:T}
T=\bar\delta_{\mathrm L}-\bar\delta_{\mathrm D}
\end{equation}
Each group has four blocks. Hold-first blocks get zero weight in the contrast but stay in the reference set. $T$ is unbiased for $\theta$, because the balanced-baseline contrast has expectation zero under $\tau$.

The exact two-sided test concerns the \emph{global sharp null} that every common-market outcome is invariant to complete-order reassignment within its stage. Under this null the paired outcome vector is fixed. Each stage has 720 assignments, which collapse to 90 equally weighted first-action arrangements. Jointly they give 8,100 plans, equivalent to 518,400 full assignments. The $p$-value is the fraction of plans with $|T|$ at least as large as observed. Because L and D can be mirrored, the minimum two-sided $p$ is $2/8{,}100$.

The analysis plan and code of the main experiment were hash-locked locally before its first model call, and this is what ``pre-specified'' means below. The mechanism analyses were chosen after the results were known and are exploratory (Sections~\ref{sec:simcheck}, \ref{sec:evaluation} and~\ref{sec:ebenchmark}; Appendices~\ref{app:audit} and~\ref{app:planned}). Table~\ref{tab:components} in Appendix~\ref{app:expe} records the status of each analysis. The locked plans, model responses, market records and analysis code are available from the author on request (Appendix~\ref{app:materials}).

\subsection{The order reaches debt only through the models}\label{sec:simcheck}

Every market in a block shares its book, schedule, seeds and engine. In the code, the complete order enters only through two prompt lines, the menu sentence and the quantity instruction. No other part of the simulator reads the order. To confirm this, I replaced the language models with three order-blind deterministic policies and ran each of the 12 common-market books under all six orders (Appendix~\ref{app:falsify}). All 216 markets give bit-identical paths across orders, and the lever-up-first minus delever-first contrast is exactly zero. A copy of the market loop that is fed the recorded responses reproduces all 24 markets of the main experiment exactly. Order dependence in the recorded outcomes therefore enters through the models' responses.

\section{A Menu Order Moves a Market's Debt}\label{sec:executed}

\begin{figure}[t]
\centering
\includegraphics[width=\textwidth]{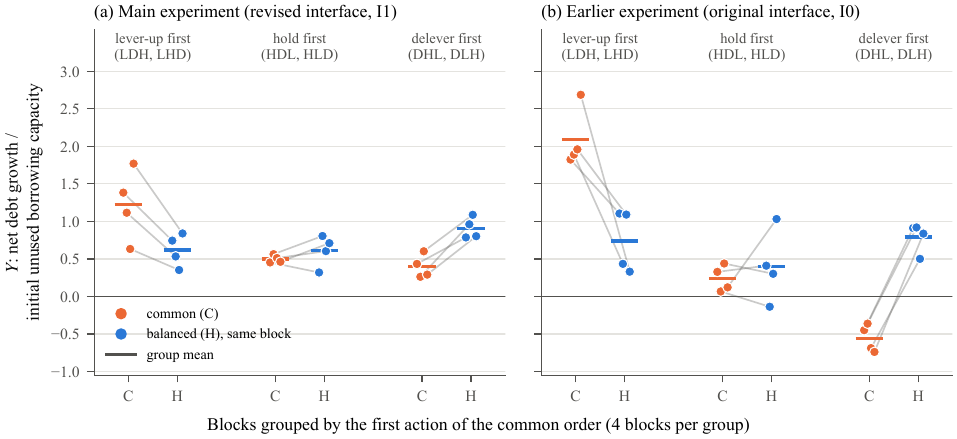}
\caption{\textbf{Executed credit by the first action of the common order.} Net debt growth $Y$ is measured in units of initial unused reference borrowing capacity $K_0$ (equation~\eqref{eq:y}). Lines join the common (C, orange) and balanced (H, blue) markets of each block. Bars mark means over four blocks and pool the two complete orders named above each group. (a)~The main experiment, revised interface I1. (b)~The earlier experiment, original interface I0. Outcomes by first action are pre-specified descriptive outputs. The panels share a scale. The two experiments differ in several ways (Appendix~\ref{app:history}).}
\label{fig:paired}
\end{figure}

A shared order moved the debt of whole markets. Figure~\ref{fig:paired}a shows all 24 markets of the main experiment, and Table~\ref{tab:emain} reports the test.

\begin{table}[t]
\centering
\caption{\textbf{Executed credit and the pre-specified test in the main experiment.} Panel A reports mean net debt growth $Y$ by the first action of the common order. Panel B reports $T$, its exact two-sided test, the influence check and sharp-null reference summaries. Panel C summarizes all 12 blocks.}
\label{tab:emain}
\small
\begin{threeparttable}
\begin{tabular}{lrrrrr}
\toprule
 & & \multicolumn{3}{c}{Mean $Y$} & \\
\cmidrule(lr){3-5}
Blocks & $n$ & Common (C) & Balanced (H) & C $-$ H & C $>$ H \\
\midrule
\multicolumn{6}{l}{\emph{A. By the action the common order lists first}} \\
Lever-up first & 4 & 1.225 & 0.617 & 0.608 & 4/4 \\
Hold first & 4 & 0.497 & 0.609 & $-$0.112 & 1/4 \\
Delever first & 4 & 0.397 & 0.909 & $-$0.513 & 0/4 \\
\midrule
\multicolumn{6}{l}{\emph{B. Pre-specified statistic and exact test}} \\
$T$: lever-up first minus delever first & & 0.828 & $-$0.293 & 1.121 & \\
Exact two-sided $p$ & & & & 2/8,100 & \\
Leave-one-block-out range of $T$ (exploratory) & & & & 1.013--1.230 & \\
SD of $T$ over the 8,100 plans (exploratory) & & & & 0.38 & \\
\midrule
\multicolumn{6}{l}{\emph{C. All blocks (descriptive)}} \\
Mean & 12 & 0.706 & 0.712 & $-$0.006 & 5/12 \\
Standard deviation & 12 & 0.47 & 0.23 & & \\
Ratio of standard deviations, C/H & & \multicolumn{2}{c}{2.05} & & \\
\bottomrule
\end{tabular}
\begin{tablenotes}[flushleft]\footnotesize
\item \textit{Note.} $Y$ is net debt growth per unit of initial unused reference capacity $K_0$ (equation~\eqref{eq:y}); 12 blocks, 24 markets, interface I1. Panel B computes $T$ as the C entry minus the H entry, retaining both components. The exact test uses 8,100 equally weighted plans and tests complete-order invariance conditional on the balanced allocation and locked inputs (Section~\ref{sec:estimand}). $T$ estimates $\theta$. The leave-one-block-out range recomputes $T$ with each block removed in turn; the reference-set SD describes the reassigned statistic under the sharp null and is not a standard error or an interval for $\theta$. Panels A and C are pre-specified descriptive outputs. Differences are computed before rounding.
\end{tablenotes}
\end{threeparttable}

\end{table}

\subsection{The order changes net debt}\label{sec:etest}

The pre-specified test rejects complete-order invariance ($p = 2/8{,}100$). This is the smallest two-sided value the design allows. Only the observed plan and its lever-up/delever mirror produce $|T|$ this large (Appendix Figure~\ref{fig:refdist}). Rejection means that the order changed the net debt of at least one common market. The estimate of $T = 1.121$ units of $K_0$ gives the direction and size, and its sign is the one predicted. $T$ is unbiased for the order contrast $\theta$ (equations~\eqref{eq:theta}--\eqref{eq:T}). It measures the total effect of the order with the model mix and market inputs fixed. That total runs through requests, execution, price interaction and later decisions.

\subsection{Size and direction}\label{sec:reading}

Each common market ended on the side of its first action relative to its balanced partner. Every lever-up-first market ended above its partner (4 of 4), by 0.608 $K_0$ on average. Every delever-first market ended below its partner, by 0.513 on average. $T$ is the difference between these two paired means. One of four hold-first markets ended above its partner.

In levels, lever-up-first markets increased net debt by 1.225 $K_0$ on average, against 0.397 in delever-first markets. This gap of 0.828 equals 3,767 USDC per market (5,559 against 1,792). The remaining 0.293 of $T$ comes from the balanced baselines assigned to the two groups (0.909 against 0.617). This component averages to zero over $\tau$ given $\sigma$. It remains part of the pre-specified estimate.

No single block drives the result. $T$ stays between 1.01 and 1.23 when any one block is omitted. The sharp-null reference distribution has SD 0.38 and a central 95 percent range of $\pm$0.74, so reassigning orders rarely produces an estimate this large (Table~\ref{tab:emain}, Panel B; Appendix Figure~\ref{fig:refdist}).

The order decides how much debt grows, not whether it grows. All 24 markets increased debt, including the delever-first markets (0.26--0.60 $K_0$). Borrowing exceeded repayment in every market (Section~\ref{sec:execution}).

The shock also cuts both ways, so it does not show up in averages. The pooled means are nearly equal, at 0.706 for common and 0.712 for balanced markets. Common outcomes are about twice as dispersed (SD 0.47 against 0.23; ratio 2.05). This comparison is descriptive, and it fits the pattern above. Markets that share an order end on opposite sides of their partners, and the pooled mean averages these shifts away.

\subsection{The earlier experiment points the same way}\label{sec:bfindings}

The main experiment was built to test a pattern first seen in an earlier experiment under the original interface. That experiment's pre-specified Pitman--Morgan test rejected joint C/H exchangeability over 4,096 within-block label swaps (one-sided $p = 11/4{,}096$). Its directional pattern was found afterward. There the first-action groups separate sharply (Figure~\ref{fig:paired}b). Delever-first common markets reduced debt by 0.36--0.74 $K_0$, below every balanced market. Lever-up-first common markets added 1.82--2.69 $K_0$, above every balanced market. In the main experiment the separation is narrower, as the lever-up-first range (0.63--1.77) overlaps the balanced range (0.32--1.09). The direction is the same, and now it passes a test locked in advance. Appendix~\ref{app:expb} reports the earlier experiment in full, and Appendix~\ref{app:history} compares the two.

\section{Why the Collateral Cap Does Not Undo the Shock}\label{sec:transmission}

A shock to choices reaches a balance sheet only if it survives execution. This section follows the order from the moment of choice to the ledger. The order moves choices, and one model family makes most of them. Debt then adjusts on two margins. The market's rules screen one margin and leave the other almost untouched.

\subsection{The order moves choices}\label{sec:requests}

The frozen-state interface test isolates the moment of choice. It presents 1,296 single decisions in 12 fixed states, paired across the two interfaces (Appendix~\ref{app:expd}). Under the revised interface, listing lever-up first rather than last raises positive lever-up requests by 15.3 percentage points (95 percent state-cluster bootstrap interval 9.3--21.3; Table~\ref{tab:remedy}). Order sensitivity therefore remains under the interface used in the main experiment.

Interface design changes the size of this response. In the same pre-specified test, the original interface gives 29.2 points, so the revision reduces the effect by 13.9 points (95 percent interval $-23.1$ to $-4.6$; exact sign-flip $p = 0.026$). The reduction met the advance criterion for running a market experiment under I1. It reflects both changes together, the removal of the ordered schema line and the consistent treatment of zero quantities (Section~\ref{sec:interfaces}; Appendix~\ref{app:expd}). An earlier frozen-state test found a 27.3-point effect under the original interface (95 percent interval 23.6--31.2). A decomposition of that prompt attributes the largest main effect to the order of the action values in the JSON schema example (18.5 points; Appendix~\ref{app:frozen}). Section~\ref{sec:scope} and Appendix~\ref{app:larger} report the same test on three larger configurations, model by model.

The response also varies with the borrower's position. An exploratory comparison of the 12 states describes the effect by pool-valued LTV (Appendix~\ref{app:planned}). Under the revised interface, the lever-up-first effect on borrowing requests was 22, 11 and 13 points at LTVs of 0.30, 0.50 and 0.66. A request gap thus remains in the high-LTV states, but only a third of the requested-token contrast is executable there. The delever-first effect on collateral-sale requests was 1, 15 and 17 points across the same states. Sale requests face no such limit.

\subsection{One family makes most of the credit decisions}\label{sec:participation}

Three families were deployed in equal numbers, and only two traded (Table~\ref{tab:pipeline}, Panel A). Each family made 1,800 decisions. Llama made 732 positive borrowing requests and 235 collateral-sale requests. Mistral made 154 borrowing requests and no sales. Qwen made 1,800 valid no-trade responses. All Mistral borrowing requests came under the opportunistic mandate, and Llama supplied all trading under the other two mandates. Llama bought 991 of the 1,159 executed tokens (86 percent). Equal representation thus produced concentrated credit flows.

These flows carry the effect. $T$ is linear in debt flows, so an exploratory accounting split is exact. It assigns 0.723 of $T=1.121$ to Llama borrowers' net debt changes, 0.398 to Mistral and zero to Qwen. The frozen-state test helps explain the zero. The lever-up-first effect fell from 21 points under I0 to zero under I1 for Qwen, from 46 to 28 for Mistral and from 21 to 18 for Llama (Figure~\ref{fig:remedy}; Appendix~\ref{app:expd}). Qwen's response disappeared under the revised interface, and the two families that trade stayed sensitive to the order.

\begin{table}[t]
\centering
\caption{\textbf{Requests, execution and debt flows in the main experiment.} I1; 24 markets, 5,400 decisions. Panel A is pre-specified descriptive. Debt-reducing sales ($\dagger$) and Panel B are exploratory accounting. Positive requests and \emph{any execution} require at least one token. No trade includes 51 active-action responses with zero quantity. Invalid responses execute as no trade and stay in the denominators. Panel B reports borrowing minus repayment in units of $K_0$, averaged over four markets per group. Balanced markets are grouped by their paired common order. Differences precede rounding. Market-level flows are in Appendix~\ref{app:expe}.}
\label{tab:pipeline}
\footnotesize
\setlength{\tabcolsep}{3.5pt}
\textit{Panel A. Requests, participation and execution}\\[2pt]
\begin{tabular}{lrrrrrrrr}
\toprule
 & \multicolumn{3}{c}{Lever-up (borrowing)} & \multicolumn{3}{c}{Delever (collateral sale)} & & \\
\cmidrule(lr){2-4}\cmidrule(lr){5-7}
 & Positive & With any & Tokens & Positive & With any & Reduced & No & \\
 & requests & execution & req.\ / exec. & requests & execution & debt$^{\dagger}$ & trade & Invalid \\
\midrule
\textit{Model family} & & & & & & & & \\
Llama & 732 & 563 & 1,360 / 991 & 235 & 230 & 216 & 797 & 36 \\
Mistral & 154 & 73 & 404 / 168 & 0 & 0 & 0 & 1,645 & 1 \\
Qwen & 0 & 0 & 0 / 0 & 0 & 0 & 0 & 1,800 & 0 \\
\addlinespace
\textit{Mandate} & & & & & & & & \\
Opportunistic & 509 & 310 & 1,019 / 547 & 51 & 51 & 51 & 1,227 & 13 \\
Neutral & 264 & 222 & 540 / 432 & 74 & 72 & 70 & 1,450 & 12 \\
Capital preservation & 113 & 104 & 205 / 180 & 110 & 107 & 95 & 1,565 & 12 \\
\addlinespace
Total & 886 & 636 & 1,764 / 1,159 & 235 & 230 & 216 & 4,242 & 37 \\
\bottomrule
\end{tabular}

\vspace{0.8em}
\textit{Panel B. Borrowing and repayment per unit of $K_0$ (exploratory)}\\[2pt]
\begin{tabular}{llrrr}
\toprule
 & First action of & & & Net debt \\
Markets & the common order & Borrowing & Repayment & change ($Y$) \\
\midrule
Common (C) & lever-up & 1.415 & 0.190 & 1.225 \\
 & hold & 0.769 & 0.272 & 0.497 \\
 & delever & 0.938 & 0.541 & 0.397 \\
\addlinespace
Balanced (H) & lever-up & 1.004 & 0.387 & 0.617 \\
 & hold & 0.926 & 0.317 & 0.609 \\
 & delever & 1.322 & 0.413 & 0.909 \\
\midrule
\multicolumn{5}{l}{\textit{Lever-up first minus delever first}} \\
Common (C) &  & 0.477 & $-$0.351 & 0.828 \\
Balanced (H) &  & $-$0.319 & $-$0.026 & $-$0.293 \\
C minus H ($T$) &  & 0.796 & $-$0.325 & 1.121 \\
\bottomrule
\end{tabular}

\end{table}

\subsection{Debt adjusts on two margins}\label{sec:execution}

Without liquidation or realized bad debt, net debt change equals borrowing minus repayment. In an exploratory accounting split (Table~\ref{tab:pipeline}, Panel B), lever-up-first markets borrowed 1.415 $K_0$ and repaid 0.190. Delever-first markets borrowed 0.938 and repaid 0.541. Their 0.828 gap is therefore 0.477 more borrowing plus 0.351 less repayment. For $T=1.121$, the same split is 0.796 more borrowing and 0.325 less repayment. Exploratory randomization tests on the 8,100 plans reject order invariance for borrowing ($p=2/8{,}100$) and, more weakly, for repayment ($p=388/8{,}100$).

The two margins meet very different execution. The borrowing cap screens requests (equation~\eqref{eq:cap}). Of 886 positive borrowing requests, 636 (71.8 percent) executed, 48 of them partially. The other 250 executed nothing. Borrowers bought 1,159 of 1,764 requested tokens (65.7 percent). Collateral sales meet no such limit. They executed in 230 of 235 requests and sold 435 of 454 offered tokens. Of these sales, 216 reduced debt. The other 14 were sales by Llama borrowers without debt, which added wallet cash.

Collateral sales are also the only way to reduce debt. Holding and wallet cash do not repay debt, and there is no interest. Even so, borrowing exceeded repayment in all 24 markets. Borrowers began with unused capacity, and purchases added collateral. The spot-valued cap also used the prices those purchases raised.

\subsection{The rules screen borrowing and not selling}\label{sec:evaluation}

\begin{table}[t]
\centering
\caption{\textbf{Fixed-request decomposition of the common-market order gap.} Exploratory counterfactuals. The recorded requests of the 12 common markets of the main experiment are executed under four sets of rules, adding one rule at a time. Environment D is the recorded market.}
\label{tab:decomposition}
\footnotesize
\begin{threeparttable}
\setlength{\tabcolsep}{2.6pt}
\begin{tabular}{lrrrrrrcc}
\toprule
 & \multicolumn{3}{c}{Mean $Y$} & Change & \multicolumn{2}{c}{Gap in} & Lever-up tokens & Fully blocked \\
\cmidrule(lr){2-4}\cmidrule(lr){6-7}
Environment & L first & D first & Gap & in gap & borrowing & repayment & executed & lever-ups \\
\midrule
A. Accounting at price 100, no cap & 2.282 & 0.733 & 1.549 &  & 1.174 & $-$0.376 & 449 / 233 & 0 / 0 \\
B. + reference-valued 70\% cap & 1.124 & 0.386 & 0.738 & $-$0.811 & 0.385 & $-$0.353 & 239 / 167 & 76 / 35 \\
C. + AMM prices and arbitrage & 1.061 & 0.374 & 0.688 & $-$0.051 & 0.336 & $-$0.351 & 222 / 163 & 80 / 38 \\
D. Pool-valued cap (recorded) & 1.225 & 0.397 & 0.828 & 0.141 & 0.477 & $-$0.351 & 250 / 167 & 73 / 36 \\
\bottomrule
\end{tabular}
\begin{tablenotes}[flushleft]\footnotesize
\item \textit{Note.} L first and D first are the four lever-up-first and four delever-first common markets. Each market's recorded requests (action, quantity and position in the round) are executed in every environment. Agents do not observe these states or decide again. $Y$ and flows are means in units of $K_0$. The gap is lever-up first minus delever first. It equals the borrowing gap minus the repayment gap, since no liquidation occurs. The change is the gap minus the previous row's gap, from unrounded values. The total change is $-$0.721. Token and request counts are totals over four markets, given as L / D. Requested lever-up tokens are 449 / 233 and lever-up requests 187 / 153 in every environment.
\end{tablenotes}
\end{threeparttable}

\end{table}

To see what execution does, I re-execute each common market's recorded requests under four sets of rules (Table~\ref{tab:decomposition}). This exercise is exploratory. It holds actions, quantities and positions in the round fixed, so it gives an accounting of the recorded requests. Environment A executes every borrowing request in full at the reference price of 100, with no cap. B adds the 70 percent cap valued at 100. C adds the AMM's prices, fee and arbitrage and keeps the reference-valued cap. D values the cap at the post-trade pool price, as in equation~\eqref{eq:cap}, and reproduces every recorded decision, round and final book. No liquidation occurs in any of the four.

The gap in requests is almost twice the gap on the ledger. The requests alone imply a common-market gap of 1.549 capacity units. Lever-up-first markets requested 449 lever-up tokens against 233, and they sold less collateral. The cap is the largest step. It cuts the gap by 0.811, to 0.738, by executing 239 of the 449 requested tokens in lever-up-first markets and 167 of 233 in delever-first markets. AMM pricing cuts the gap by a further 0.051. Valuing collateral at the pool price raises it by 0.141, to 0.828. These steps act mainly on borrowing, as the repayment gap stays between $-0.376$ and $-0.351$. Appendix~\ref{app:decomp} gives the alternative order of the steps. The same chain for $T$, which includes the balanced baseline, runs from 1.905 in A to 1.121 in D.

The two margins therefore pass through differently. From A to D, 41 percent of the common-market borrowing difference survives, against 93 percent of the repayment difference. These are ratios along one fixed-request path. The repayment margin carries 24 percent of the gap in A and 42 percent in D. The asymmetry follows from the rules. The cap limits new borrowing, but no rule makes a borrower sell collateral before liquidation. A difference in how much borrowers sell therefore reaches the ledger almost unscreened. Lever-up-first markets executed 8.75 sale tokens per market, against 30.25 in delever-first markets.

\subsection{Borrowing uses up capacity, and valuation returns part of it}\label{sec:headroom}

Two features of the cap explain the borrowing side. First, a purchase uses up the buyer's own room. Write borrower $i$'s remaining borrowing room at pool price $p$ as $H_i=0.70\,c_i\,p-d_i$. A purchase of $q$ tokens raises the price to $p'$ and changes the buyer's room by
\begin{equation}\label{eq:headroom}
\Delta H_i=\underbrace{0.70\,c_i\,(p'-p)}_{\text{own collateral revalued}}+\underbrace{0.70\,q\,p'}_{\text{new collateral}}-\underbrace{\beta(q)}_{\text{purchase cost}}.
\end{equation}
The purchase cost exceeds the first two terms unless the buyer holds more than 21 to 22 percent of the pool's tokens (Appendix~\ref{app:headroom}). The largest buyer in the main experiment held 3.0 percent. All 636 executed purchases raised the pool price and lowered the buyer's own room. Summed over all borrowers, room fell after 618 of them. Borrowing therefore uses up capacity, and the cap cuts hardest where requests are largest.

Second, a purchase has price impact, and under the pool-valued cap this reaches other borrowers. It raises every other borrower's room by $0.70\,c_j\,(p'-p)$, and it raises the price that a later buyer pays. Under the pool-valued cap, the first effect wins for any purchase that was already feasible. If the later borrower's position is unchanged and no sale, arbitrage or liquidation intervenes, that purchase stays feasible with more slack. A purchase that was blocked can become feasible. Under a reference-valued cap, the later buyer pays the higher price but gains no revaluation, so its feasible purchases can only shrink.

This is why pool valuation returns part of the screened borrowing, and the return is largest where purchases cluster. Net debt under D exceeds net debt under C by 0.164 $K_0$ in the lever-up-first group and by 0.023 in the delever-first group. Under D, 16 borrowers in lever-up-first markets end above 70 percent of reference collateral value. Under C, none do. Pool valuation raises the gap only from 0.688 to 0.828, but it creates all of these positions. These borrowers comply with the cap and still carry high leverage at an outside price. The next section shows why that matters.

\section{Debt Gaps Become Larger Exposure Gaps}\label{sec:losses}

A loan book matters most when prices fall. None of the 48 markets had a liquidation or realized bad debt (Table~\ref{tab:losses}, Panel A). I therefore apply specified shocks to the recorded books, so every exposure in this section is a scenario value. The main result is exploratory. In proportional terms, the order moves scenario exposure more than debt. A threshold in the liquidation formula explains why.

\subsection{How a shock turns a book into a loss}\label{sec:books}\label{sec:realized}\label{sec:formula}

I impose an instantaneous terminal shock that moves the outside collateral price to $F=100(1-s)$. Arbitrage moves the pool price to $F$, and borrowers stay passive. Before liquidation, collateral may already fall short of debt. This \emph{pre-liquidation deficiency} is
\begin{equation}\label{eq:u0}
U_0(F) = \textstyle\sum_i \max(d_i - F c_i,\, 0).
\end{equation}
A \emph{scenario write-off}, $W(F)$, occurs when liquidation exhausts collateral. \emph{Remaining deficiency} is uncovered debt on accounts still open when liquidation stops. Total \emph{creditor exposure} adds these two disjoint parts.
\begin{equation}\label{eq:exposure}
U(F) = W(F) + \textstyle\sum_{i\,\text{open}} \max(d_i^{\text{final}} - F c_i^{\text{final}},\; 0).
\end{equation}

\begin{table}[tp]
\centering
\caption{\textbf{Realized losses and conditional creditor exposure.} Panel A reports run totals and the main experiment's pre-specified descriptive benchmark ($F=70$, 8 percent bonus, completed liquidation and unlimited funding). $B$ includes $U_0$. Panel B applies three deterministic rules with profit-seeking liquidators to the main experiment's books (exploratory). The 30 percent no-feedback row equals Panel A's $B(70)$. Means use 12 markets per arm; exposure/debt uses pooled arm debt. Equations~\eqref{eq:u0}--\eqref{eq:bonus} define the measures. Appendix Table~\ref{tab:lossesorig} reports the earlier experiment's books.}
\label{tab:losses}
\footnotesize
\setlength{\tabcolsep}{3.5pt}
\textit{Panel A. Realized losses (both market experiments) and the main experiment's fixed-price benchmark}\\[2pt]
\begin{tabular}{lrr}
\toprule
 & Common (C) & Balanced (H) \\
\midrule
\multicolumn{3}{l}{\textit{Liquidations / realized bad debt during the run (USDC)}} \\
Earlier experiment & 0 / 0 & 0 / 0 \\
Main experiment & 0 / 0 & 0 / 0 \\
\addlinespace
\multicolumn{3}{l}{\textit{Fixed-price benchmark of the main experiment, $F = 70$ (pre-specified)}} \\
Scenario write-off $B(70)$, USDC per market & 323 & 282 \\
\quad per unit of $K_0$ & 0.0705 & 0.0607 \\
Pre-liquidation deficiency $U_0(70)$, USDC per market & 57 & 27 \\
\quad per unit of $K_0$ & 0.0128 & 0.0059 \\
Remaining deficiency after liquidation & 0 & 0 \\
\bottomrule
\end{tabular}

\vspace{0.8em}
\textit{Panel B. Three liquidation rules on the main experiment's books (exploratory); USDC per market}\\[2pt]
\begin{tabular}{llrrrrrr}
\toprule
 & & \multicolumn{2}{c}{Creditor} & \multicolumn{2}{c}{of which} & \multicolumn{2}{c}{Exposure /} \\
 & & \multicolumn{2}{c}{exposure} & \multicolumn{2}{c}{remaining} & \multicolumn{2}{c}{pre-shock} \\
 & & \multicolumn{2}{c}{$U(F)$} & \multicolumn{2}{c}{deficiency} & \multicolumn{2}{c}{debt (\%)} \\
\cmidrule(lr){3-4}\cmidrule(lr){5-6}\cmidrule(lr){7-8}
Rule & Shock & C & H & C & H & C & H \\
\midrule
No feedback & 10\% & 0 & 0 & 0 & 0 & 0.0 & 0.0 \\
 & 20\% & 0 & 0 & 0 & 0 & 0.0 & 0.0 \\
 & 30\% & 323 & 282 & 0 & 0 & 3.6 & 3.1 \\
\addlinespace
External trigger, pool sale & 10\% & 0 & 0 & 0 & 0 & 0.0 & 0.0 \\
 & 20\% & 0 & 0 & 0 & 0 & 0.0 & 0.0 \\
 & 30\% & 215 & 178 & 174 & 157 & 2.4 & 2.0 \\
\addlinespace
Pool trigger, pool sale & 10\% & 0 & 0 & 0 & 0 & 0.0 & 0.0 \\
 & 20\% & 134 & 28 & 0 & 0 & 1.5 & 0.3 \\
 & 30\% & 586 & 507 & 0 & 0 & 6.5 & 5.6 \\
\bottomrule
\end{tabular}

\end{table}

Liquidation sets the write-off. Under an external price, an account can be liquidated when $d_i>0.80Fc_i$. With an 8 percent bonus, repayment of $k d_i$ seizes $1.08 k d_i/F$ units of collateral. Thus
\begin{equation}\label{eq:invariant}
d_i' - \frac{F c_i'}{1.08} \;=\; (1-k)\,d_i - \frac{F c_i}{1.08} + k\,d_i \;=\; d_i - \frac{F c_i}{1.08}.
\end{equation}
Positive values exhaust collateral and become write-offs. Negative values leave covered loans, and equality gives zero write-off. With completed liquidation,
\begin{equation}\label{eq:bonus}
B(F) = \textstyle\sum_i \max\!\left(d_i - \dfrac{F c_i}{1.08},\; 0\right).
\end{equation}
Under no feedback $W(F)$ equals $B(F)$, which already includes pre-liquidation deficiency. When repayment and seizure happen at $F$ and liquidation completes, no open account remains deficient and $U(F)=B(F)$. Appendix~\ref{app:stress} gives the completion conditions and the engine check.

\begin{figure}[tp]
\centering
\includegraphics[width=\textwidth]{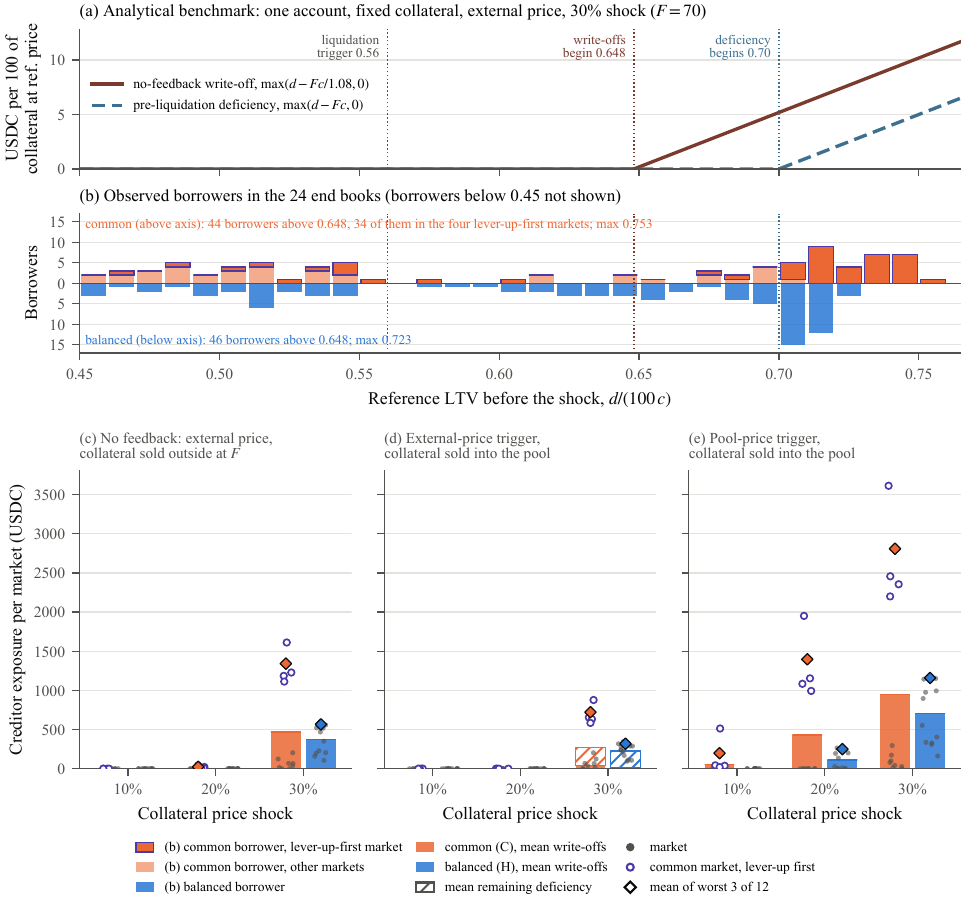}
\caption{\textbf{Liquidation thresholds and conditional creditor exposure.} (a) Fixed-price benchmark at a 30 percent shock, with fixed collateral, an 80 percent liquidation threshold and an 8 percent bonus. $B$ and $U_0$ are per 100 USDC of reference collateral. Panels (b) to (e) use the terminal books of the earlier experiment. (b) Borrower reference LTVs, with common books above and balanced books below. Darker bars identify lever-up-first common markets. (c) to (e) Deterministic exposure under three profit-seeking liquidation rules. Solid bars are write-offs and hatched bars remaining deficiency. Dots are market values, open violet circles mark lever-up-first common markets and diamonds are worst-three means. Appendix Table~\ref{tab:lossesorig} and Appendix~\ref{app:stress} give the values.}
\label{fig:stress}
\end{figure}

Write the reference LTV as $\ell_i=d_i/(100c_i)$. At shock size $s$, the thresholds are
\begin{equation}\label{eq:thresholds}
\text{liquidation if } \ell > 0.80(1-s), \qquad \text{write-off if } \ell > \frac{1-s}{1.08}, \qquad \text{deficiency if } \ell > 1-s.
\end{equation}
At a 30 percent shock the thresholds are 0.56, 0.648 and 0.70 (Figure~\ref{fig:stress}a). The bonus opens the gap between the last two. A loan at 70 percent of reference collateral value has no deficiency at $F=70$. Liquidation still writes off 5.2 USDC per 100 USDC of reference collateral value, because the bonus consumes collateral that would otherwise cover the creditor. A 70 percent reference-valued cap therefore still permits write-offs under this shock.

\subsection{The order moves exposure more than debt}\label{sec:ebenchmark}

The main experiment pre-specified one benchmark. It fixes $F=70$ and completes liquidation with an 8 percent bonus and unlimited funding, so write-offs follow equation~\eqref{eq:bonus} and remaining deficiency is zero. The complete order was randomized across common markets, so their books can be compared by first action. This comparison is exploratory.

Lever-up-first markets ended with 58 percent more debt than delever-first markets (11,435 against 7,218 USDC per market). Their aggregate reference LTV was 0.559 against 0.470, and the groups' ranges barely touch (0.489--0.620 against 0.453--0.489). Sixteen of their borrowers ended above a reference LTV of 0.70, against 2. Benchmark write-offs were 630 USDC per market against 151, a ratio of 4.2. These differences have the same sign in both stages. In lever-up-first blocks, write-offs in the common market were 405 USDC higher than in the paired balanced market. In delever-first blocks they were 202 USDC lower. Exploratory randomization tests on the 8,100 plans reject order invariance of benchmark write-offs ($p=2/8{,}100$), pre-liquidation deficiency ($p=12/8{,}100$), aggregate reference LTV ($p=4/8{,}100$) and the number of borrowers above the write-off threshold ($p=36/8{,}100$; Appendix~\ref{app:planned}).

Exposure responds more than debt because of the threshold. Every market's aggregate reference LTV was at most 0.620, below 0.648. Write-offs therefore come only from individual borrowers above the threshold. For borrowers with positive collateral,
\[
B(F) = \textstyle\sum_i 100\,c_i \max\!\left(\ell_i - \dfrac{1-s}{1.08},\; 0\right) \;\ge\; 100\,C \max\!\left(\bar\ell - \dfrac{1-s}{1.08},\; 0\right),
\]
where $C=\sum_i c_i$ and $\bar\ell=\sum_i d_i/(100C)$ is aggregate reference LTV. With collateral fixed, equal borrower LTVs attain this bound, so aggregate coverage can hide borrowers above the threshold. In the main experiment, an equal-LTV allocation of the same total debt removes write-offs in all 24 markets under this benchmark. Scaling the recorded distribution down to the market's initial aggregate LTV also removes them. Write-offs therefore need both a high level of debt and an uneven distribution. With a pool-price trigger, the equal-LTV allocation still leaves exposure in two lever-up-first markets at a 30 percent shock (Appendix~\ref{app:eexpl}).

The exposure gap tracks the level of debt, and dispersion shows no consistent rise. The distribution of debt was already uneven, and higher debt moved more borrowers across the threshold. The Gini coefficient of borrower LTV is 0.171 against 0.197, the coefficient of variation 0.319 against 0.373 and the normalized HHI of debt shares 0.033 against 0.035. Each measure is higher in lever-up-first markets in the first stage and lower in the second, and the groups' ranges overlap. Appendix~\ref{app:alloc} reports all 17 level, threshold, dispersion and exposure metrics for both market experiments.

The liquidation bonus accounts for most of the benchmark write-off. Write-offs averaged 323 USDC per common market and 282 per balanced market (0.0705 and 0.0607 of $K_0$; Table~\ref{tab:losses}, Panel A). Pre-liquidation deficiency averaged 57 and 27 USDC. The bonus component, $B-U_0$, accounts for 266 of 323 USDC and 255 of 282. Loans that are covered before liquidation can therefore produce write-offs during liquidation. These averages also show where the risk lies. The descriptive gap between common and balanced markets is only 41 USDC per market (0.0098 $K_0$). The large exposure gap lies between orders.

\subsection{Liquidation rules change the exposure of the same books}\label{sec:rules}\label{sec:arms}

The same books carry different exposure under different liquidation rules (Table~\ref{tab:losses}, Panel B; Figure~\ref{fig:stress}c--e for the earlier experiment's books). I evaluate them on a table of 1,800 deterministic cells. It has 19 rules at three shocks on the reference base (1,368 cells) and six core rules at three shocks on the pool base (432 cells). The grid was designed for the earlier experiment's 24 terminal books and frozen after the market outcomes were known but before any stress output. Its application to the main experiment's books is exploratory.

Take those books at a 30 percent shock. An external-price trigger with pool settlement gives 215 and 178 USDC per common and balanced market. Of this, 174 and 157 are remaining deficiency on accounts that profit-seeking liquidators leave open. Write-offs alone would miss most uncovered claims. A pool-price trigger gives 586 and 507 USDC, because liquidation sales lower the price that triggers further liquidations. It is the only headline rule with exposure at a 20 percent shock (134 and 28 USDC). No rule produces exposure at a 10 percent shock. The earlier experiment's books carry more leverage and show the same ranking of rules with larger values (Appendix~\ref{app:stress}). There, a reference-valued cap removes positions above 70 percent of reference value and leaves bonus-related write-offs and liquidation feedback (Appendix~\ref{app:replay}).

Across rules, I find no consistent gap between common and balanced books on average. Three of the 19 rules give lower common-market mean exposure at 30 percent. Exploratory worst-three common/balanced ratios range from 1.31 to 2.15. On the earlier experiment's books, the paired sign-flip calculation gives $p=0.61$ for the no-feedback rule (Appendix~\ref{app:stress}). As with debt, the risk depends on which order a market shares.

\section{Implications}\label{sec:implications}

\subsection{A shared interface is a common input}\label{sec:assessment}

A shared menu order acts as a common shock to debt adjustment. It carries no payoff-relevant information, yet it changed executed debt. Every borrower in a common market saw the same order, so its effect did not average out across borrowers. Each common market ended on the side of its first-listed action relative to the balanced market of its block (Section~\ref{sec:reading}).

Supervisory discussions emphasize common models, data and providers \citep{fsb2024ai,danielsson2022ai}. A shared decision interface belongs on that list. With small open-weight models in this environment, it reached executed credit. The paired frozen-state test shows that interface design can reduce request-level order sensitivity. The market evidence shows that the remaining sensitivity can survive execution. Exploratory counterfactuals suggest that it survives most fully on the margin that borrowing constraints do not police. An interface is therefore a design choice with credit consequences.

\subsection{What collateral rules police}\label{sec:policy}

The collateral constraint screens the two margins of this shock differently. It limits new borrowing, and each purchase uses up the buyer's own borrowing room. It does not require anyone to sell collateral. In the exploratory fixed-request counterfactuals, the rules cut the borrowing difference by more than half and leave the repayment difference almost intact (Section~\ref{sec:evaluation}). A difference in deleveraging therefore reaches the ledger more fully than a difference in borrowing. Frozen states show the same asymmetry in requests. At a pool-valued LTV of 0.66, a request gap for borrowing remained but mostly could not execute. The delever-first effect on sale requests was larger at LTVs of 0.50 and 0.66 than at 0.30 (Appendix~\ref{app:planned}).

Valuation adds a spillover. The pool-valued cap implements collateral-price feedback \citep{kiyotaki1997credit,geanakoplos2010leverage}. A purchase raises the price used to value every borrower's collateral. The buyer pays for the purchase and loses borrowing room, while other borrowers gain room (equation~\eqref{eq:headroom}). Cap compliance can therefore coexist with low coverage at an outside price. Price-dependent constraints of this kind create collateral externalities \citep{davila2018pecuniary}. I measure this spillover as the change in other borrowers' borrowing room.

Liquidation design then decides how a higher level of debt becomes a scenario loss. Scenario write-offs arose only where individual borrowers crossed the threshold, and higher leverage moved more borrowers across. Lever-up-first markets had 1.58 times the debt of delever-first markets and 4.2 times the benchmark write-offs. The 8 percent liquidation bonus lowers the write-off threshold below the deficiency threshold, and it accounts for most benchmark write-off (equations~\eqref{eq:invariant}--\eqref{eq:bonus}). Pool-price triggers and slow replenishment add exposure through liquidation sales \citep{lehar2022fragility}. Profit-seeking liquidators can leave uncovered loans open. Borrower liquidation penalties are a different outcome \citep{chiu2026defilending}. Fixed-discount and auction designs create different liquidation incentives \citep{qin2021liquidations,tian2025liquidation}.

\subsection{Who acts decides where the order matters}\label{sec:scope}

Deploying several models does not guarantee several kinds of decisions. Three model families were deployed, yet Llama bought 86 percent of the tokens purchased on credit and made all collateral sales. Qwen always chose no trade, and Mistral borrowed only under the opportunistic mandate. The diversity of credit decisions was therefore lower than the diversity of deployed models. An assessment of concentration should count who acts and not only who is present.

Who acts also shapes the evidence on larger models. In the static, newsless states of the interface test, three larger configurations traded little or not at all (Appendix~\ref{app:larger}). Qwen 2.5 14B showed a lever-up-first effect of 15.3 points under I0 (95 percent interval 6.9--23.6) and chose no trade in every I1 cell. Llama 3.2 90B Vision and Gemini 4 Argon chose no trade in every cell that returned a response (399 and 427 of 432 cells), so the order had no request to move in these states.

In a market with price-relevant news, these models trade in news rounds and a shared order moves executed debt ($T=0.33$, $p=162/8{,}100$; Appendix~\ref{app:news}). In exploratory splits, almost all of this estimate comes from news rounds and most of it from Llama 90B. The small families in the same news market give $T=1.33$ ($p=2/8{,}100$). A shared order reaches credit through the borrowers who act, so an assessment of an interface should start from who trades and when.

\subsection{Conclusion}\label{sec:conclusion}

A shared interface feature that carries no payoff-relevant information changed the credit that small open-weight LLM borrowers executed in a simulated lending market. An exact randomization test rejects complete-order invariance, and the pre-specified estimate of the lever-up-first versus delever-first contrast is $T=1.12$ capacity units. Exploratory sharp-null tests also reject order invariance of borrowing and of repayment. Exploratory counterfactuals on fixed requests show that the collateral cap screened the borrowing difference but not the repayment difference. They also show that pool-price valuation returned part of the screened borrowing. In an exploratory scenario comparison, the surviving rise in leverage moved more borrowers above a write-off threshold, so scenario exposure changed proportionally more than debt.

To evaluate a shared interface, one has to follow its effect past requests into the margins that constraints do and do not police. The same approach can test other shared features of AI decision systems, such as response formats, defaults and feedback rules. How borrowers adapt to different financial rules is the natural next question.

\bibliographystyle{elsarticle-harv}
\bibliography{references}

\appendix
\renewcommand{\thetable}{A\arabic{table}}
\renewcommand{\thefigure}{A\arabic{figure}}
\renewcommand{\theequation}{A\arabic{equation}}
\setcounter{table}{0}
\setcounter{figure}{0}
\setcounter{equation}{0}
\renewcommand{\theHequation}{A\arabic{equation}}
\renewcommand{\theHtable}{A\arabic{table}}
\renewcommand{\theHfigure}{A\arabic{figure}}

\section*{Appendices}
\addcontentsline{toc}{section}{Appendices}

\section{Frozen-state experiments v1--v3}\label{app:frozen}

\paragraph{Design.} Each decision shows a fixed state with the borrower's collateral, debt, LTV and wallet, the pool price and ``No news''. There is no reference price and no market dynamics. All three experiments use I0 without feedback (Section~\ref{sec:interfaces}). They use the three model families and the three mandates (opportunistic, neutral and capital preservation). Prompt variants share a fixed seed within each (state, model) cell. v1 (144 decisions) is the discovery sample and is excluded from every v2 estimate.

v2 is the pre-specified frozen-state test (1,296 decisions). It was specified and hashed before it ran. On new seeds, it tests two contrasts in 12 main states. One moves lever-up from first to last. The other changes the mandate from neutral to opportunistic. The order contrast compares lever-up, delever, hold with its reverse. The main states cross a pool-valued LTV of 30, 50 or 66 percent with a pool price of 95 or 105 and a wallet of 150 or 1,000 USDC. Each has two independent replicates and two wordings. v2 also re-sends the 144 v1 calls with identical seeds. It adds 288 calls in eight held-out states whose values all lie outside the main grid.

v3 is the decomposition (648 decisions). It uses six of the main states. It crosses a $2^3$ factorial over the orders of the menu sentence, the JSON schema example and the quantity line with all six orders of the three actions. Its specification, including its primary contrasts and decision rules, was hashed before the first call. I use its results descriptively.

Identical seeds reproduced every v1 decision in v2 (144 of 144 actions and quantities; reason text byte-identical in 142).

\paragraph{Inference.} Effects are averaged within states over models, mandates, the other presentation factor and replicates. The two v2 replicates agree on 92 percent of decisions. Bootstrap intervals resample states in the designed grid. Sign-flip tests enumerate all $2^{12}$ (main), $2^8$ (held-out) or $2^6$ (v3) sign patterns. They are exact if the state effects are jointly invariant to independent sign reversals under the null. The smallest attainable $p$ is therefore $2/4{,}096$ with 12 states and $2/64$ with six.

\paragraph{The pre-specified test (v2).} Moving lever-up from last to first raises the probability of a valid lever-up request by 27.3 percentage points (95 percent bootstrap interval over the 12 states, 23.6 to 31.2). The state-level effect is positive in all 12 main states and in all eight held-out states. The held-out states span LTVs from 20 to 68 percent, prices from 90 to 110 and wallets from 50 to 2,000 USDC. The held-out estimate is 26.4 points (23.6 to 29.9). Switching the mandate from neutral to opportunistic raises valid lever-up requests by a similar 27.1 points (21.9 to 32.6; 22.9 points held out, 15.6 to 30.2). The exact sign-flip $p$ is 0.0005 for each main contrast and 0.0078 for each held-out contrast. Holm adjustment over the two primary contrasts gives $p = 0.001$. The wording contrast is $-1.4$ points in both sets (intervals $-3.5$ to 0.9 and $-3.5$ to 0.0; $p = 0.3438$ and 0.5000). In these frozen states, the order of a menu moves leverage requests about as much as the stated investment objective. The table of contrasts is available from the author on request.

\paragraph{Decomposition (v3).} Code each component as lever-up first or last. Let $Y(s,j,q)$ be the mean valid lever-up rate in a cell. Let the main effects $M$ be the differences averaged over the other two components, and let $I_{SJQ}$ be the three-way difference of differences. The all-components contrast then satisfies $Y(1,1,1) - Y(0,0,0) = M_S + M_J + M_Q + \tfrac14 I_{SJQ}$. The pooled estimates are $0.259 = 0.028 + 0.185 + 0.028 + 0.019$. The order of the action values in the schema example has the largest main effect, 18.5 points. The menu sentence and the quantity line have 2.8 points each. Reversing all three together gives 25.9 points. The two-way interactions cancel in this contrast but not in others. The schema effect is 0.30 when the sentence and quantity line list lever-up first and 0.11 when both list it last. A sequential attribution that switches the schema first or last would assign it between 43 and 114 percent of the total. The schema is the largest averaged component contrast, but it has no unique attribution share. Its example is text in a generic JSON-mode prompt. Family contrasts, state-bootstrap intervals and position-specific choice rates are available on request.

\paragraph{Choices by listed position, families and mandates.} Across the six congruent orders of v3, lever-up, delever and hold gain almost entirely from first position. Families differ in which components move their choices and in which actions they choose. Mandates determine the candidate actions. Ordering shifts choices among them, toward either more or less borrowing. Position-specific rates by family and mandate are available on request.

\section{The frozen-state interface test}\label{app:expd}

The test pairs the original (I0) and revised (I1) interfaces in the states of Appendix~\ref{app:frozen}. I1 removes the ordered schema list and aligns the zero-quantity rules of the prompt and the parser (Section~\ref{sec:interfaces}). Table~\ref{tab:remedy} reports the complete summary.

\paragraph{Pre-specification.} The plan fixed the design, outcomes, estimands and inference. It also fixed the decision rule and the analysis code. All of these were locally hash-locked before the first call, after the results of the earlier experiment were known. The pre-run precision assessment used the decomposition's six-order data. It assumed the same between-state spread in 12 states and uncorrelated effects across interfaces. Under these assumptions the minimum detectable reduction was 6.7 percentage points at 80 percent power. There were no deviations.

\paragraph{Interfaces.} I0 is byte-identical to the congruent six-order prompts of the earlier experiment and the decomposition. The revised interface replaces the schema line \texttt{"action": "lever\_up" | "delever" | "hold",} (whose order varies) by \texttt{"action": one of the three action names above, written exactly as shown,}. It replaces the sentence ``If you hold, set qty 0.'' by ``Not trading is always a valid choice: choose hold, or set qty 0.'' The stated quantity range of 0 to 5, the menu sentence and the quantity line are unchanged. The comparison estimates the two changes jointly.

\begin{table}[htb]
\centering
\caption{\textbf{Original and revised interfaces in frozen states.} There are 1,296 decisions, from 12 states $\times$ 3 families $\times$ 3 mandates $\times$ 6 complete action orders $\times$ 2 interfaces. They are paired by seed. Panel A gives the rate with the action listed first (two of six orders) minus the rate with it listed last (two of six). Rates are in percentage points. Token rows are mean tokens per decision. The positive lever-up rate is the pre-specified primary outcome, and the other rows are descriptive. Brackets give 95 percent state-cluster bootstrap intervals over the 12 designed states. Executable tokens use the spot-valued 70 percent cap. Panel B gives shares of 648 decisions per interface. Panel C gives the primary contrast by family (descriptive). Differences are computed before rounding.}
\label{tab:remedy}
\footnotesize
\begin{tabular}{lccc}
\toprule
 & Original interface & Revised interface & Difference \\
\midrule
\multicolumn{4}{l}{\emph{A. Effect of listing the action first rather than last}} \\
Positive lever-up request (pp) & 29.2 & 15.3 & $-$13.9 \\
 & [25.0, 33.3] & [9.3, 21.3] & [$-$23.1, $-$4.6] \\
Positive delever request (pp) & 23.6 & 11.1 & $-$12.5 \\
 & [15.3, 32.4] & [5.6, 17.1] & [$-$17.6, $-$7.4] \\
Lever-up label, any quantity (pp) & 29.2 & 16.7 & $-$12.5 \\
 & [25.0, 33.3] & [10.6, 22.7] & [$-$21.8, $-$3.2] \\
Requested lever-up tokens & 0.57 & 0.45 & $-$0.12 \\
 & [0.45, 0.72] & [0.29, 0.62] & [$-$0.29, 0.06] \\
Executable lever-up tokens & 0.52 & 0.37 & $-$0.15 \\
 & [0.39, 0.69] & [0.21, 0.55] & [$-$0.29, $-$0.01] \\
\midrule
\multicolumn{4}{l}{\emph{B. Share of all decisions (percent)}} \\
Positive lever-up request & 12.8 & 13.0 & \\
Positive delever request & 8.2 & 3.7 & \\
No trade & 79.0 & 82.9 & \\
Invalid output & 0.0 & 0.5 & \\
\midrule
\multicolumn{4}{l}{\emph{C. Lever-up effect (Panel A, first row) by model family (pp)}} \\
Llama & 20.8 & 18.1 & $-$2.8 \\
Mistral & 45.8 & 27.8 & $-$18.1 \\
Qwen & 20.8 & 0.0 & $-$20.8 \\
\bottomrule
\end{tabular}

\end{table}

\paragraph{Parsing rule and design.} A response is invalid and executes nothing if it cannot be parsed or names an unknown action. It is also invalid if it chooses lever-up or delever with a quantity that is missing, not an integer, negative or larger than 5. Hold with any quantity is a valid decision not to trade, and so is lever-up or delever with quantity 0. The action label is kept. Lever-up or delever with an integer quantity from 1 to 5 is a positive request. No response is re-asked. Re-parsing the earlier recorded responses under this rule changes the classification of 41 of the 5,400 decisions of the earlier experiment. All 41 are zero-quantity outputs, and none executed a trade. It also changes 9 decisions in v2 and 3 in v3. Positive requests, executed quantities and the recorded ledgers are unchanged.

The design fully crosses the 12 main states with the three models and the three mandates. It also crosses the six action orders and the two interfaces, with one decision per cell (1,296 decisions). One seed per state and model is shared by all 36 prompts of that cell, so each comparison of the two interfaces is paired by seed. In the six states of the decomposition, all 324 original-interface responses are identical to the earlier calls they repeat. The primary estimand is the first-listed minus last-listed rate of positive lever-up requests under each interface and its change between interfaces. First-listed orders are the two of six that put the action first. Last-listed orders are the two that put it last. Intervals are state-cluster bootstrap intervals over the 12 designed states (10,000 draws). The exact sign-flip test enumerates all $2^{12}$ sign patterns of the state-level differences.

\begin{figure}[htb]
\centering
\includegraphics[width=0.56\textwidth]{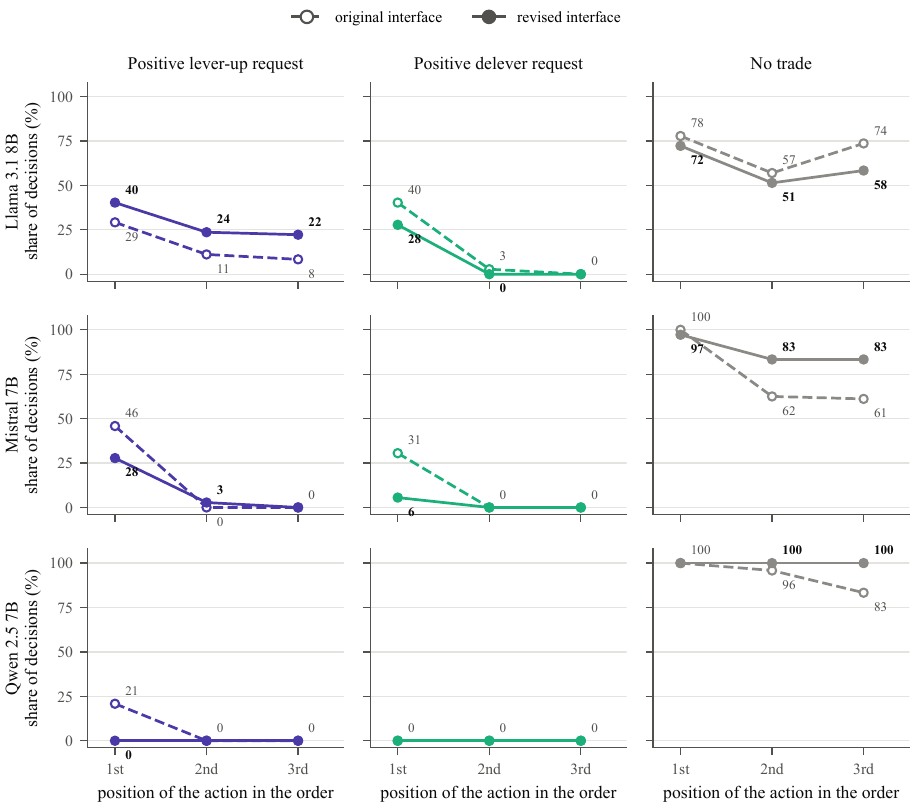}
\caption{\textbf{Frozen-state choices by action position and model family.} Shares of positive lever-up requests, positive delever requests and no trade. Dashed lines and open markers show I0. Solid lines, filled markers and bold labels show I1. Each point contains 72 decisions. Table~\ref{tab:remedy} gives the values. The rates by position are available from the author on request.}
\label{fig:remedy}
\end{figure}

\paragraph{Criterion, freeze and further results.} The revised effect of 15.3 points (9.3 to 21.3) met the pre-specified criterion for a market experiment. The criterion required at least 10 points with a lower bound above 5. Offline replay through the revised market runner reproduced all 24 ledgers of the earlier experiment, round records and final books exactly. It reclassified only 41 zero-quantity outputs. The interface, parser and feedback were then frozen for the main experiment on the same financial engine. That experiment used new books, seeds, schedules and allocation (Table~\ref{tab:campaigns}). The effect on executable lever-up tokens is 0.15 tokens per decision smaller under the revised interface (95 percent state-bootstrap interval $-0.29$ to $-0.01$). It is evaluated against each frozen state under the 70 percent cap at the pool price. Restricted to the six states of the decomposition, the lever-up effect is 30.6 points under the original interface and 14.8 under the revised one. By position (Figure~\ref{fig:remedy}), Llama requests leverage more often in every position under the revised interface. Three revised-interface outputs, all from Llama, name lever-up with a quantity of 10 and are invalid. No output under either interface pairs an active action with a quantity of zero.

\section{Design and record of the main experiment}\label{app:expe}

Table~\ref{tab:components} maps the study's evidence to its economic chain. This appendix records the main experiment's plan and its execution. Appendix~\ref{app:eexpl} reports the exploratory analyses, and Appendix~\ref{app:history} compares the two market experiments.

\begin{table}[p]
\centering
\caption{\textbf{Evidence map.} Components grouped by the link of the economic chain they address, with the statistical status and source sections of each.}
\label{tab:components}
\scriptsize
\setlength{\tabcolsep}{4pt}
\renewcommand{\arraystretch}{0.96}
\begin{tabularx}{\linewidth}{@{}>{\raggedright\arraybackslash\hsize=0.85\hsize}X>{\raggedright\arraybackslash\hsize=1.1\hsize}X>{\raggedright\arraybackslash\hsize=0.65\hsize}X>{\raggedright\arraybackslash\hsize=1.4\hsize}X>{\raggedright\arraybackslash}p{1.65cm}@{}}
\toprule
Question & Component & Setting & Evidence status & Where \\
\midrule
\multicolumn{5}{@{}l}{\emph{Main question. Does the complete action order of a shared menu reach executed credit?}} \\
Under the revised interface & Main experiment; exact sharp-null test; $T$ estimates $\theta$ & I1; 24 markets; new books & Pre-specified test. Endpoints, group means, paired signs and arm SDs pre-specified descriptive; leave-one-out $T$ and reference set descriptive & \S\ref{sec:etest}--\ref{sec:reading}; App.~\ref{app:expe} \\
Under the original interface & Earlier experiment & I0; 24 markets & Pre-specified test of C/H exchangeability (one-sided $p = 11/4{,}096$); direction analysis post hoc & \S\ref{sec:bfindings}; App.~\ref{app:expb} \\
Do the market experiments differ at market level? & Historical comparison of the two market experiments & I0 vs.\ I1; different books, seeds and draws & Descriptive & App.~\ref{app:history} \\
Does the simulator produce an order effect? & Order-blind policies, six orders & I1 engine; no model calls & Exploratory check; contrast exactly zero & \S\ref{sec:simcheck}; App.~\ref{app:falsify} \\
\midrule
\multicolumn{5}{@{}l}{\emph{Link 1. From the order to requests}} \\
Does the order move requests? & Pre-specified frozen-state test (v2) & I0; frozen; 1,296 decisions & Pre-specified test (lever-up first vs.\ last, 27.3 pp) & \S\ref{sec:requests}; App.~\ref{app:frozen} \\
Which prompt line carries it? & Decomposition (v3) & I0; frozen; 648 decisions & Descriptive; specification hashed before its first call & \S\ref{sec:requests}; App.~\ref{app:frozen} \\
Does it survive I1? & Frozen-state interface test & I0 vs.\ I1; frozen; 1,296 decisions & Pre-specified test (I1, 15.3 pp) & \S\ref{sec:requests}; App.~\ref{app:expd} \\
\midrule
\multicolumn{5}{@{}l}{\emph{Link 2. From requests to executed positions}} \\
Who trades, and what executes? & Request pipeline; participation & I1; 24 markets & Pre-specified descriptive & \S\ref{sec:participation}--\ref{sec:execution} \\
What moves net debt? & Borrowing and repayment flows; debt-reducing sales; family accounting split of $T$ & I1; decision records & Exploratory (accounting split) & \S\ref{sec:participation}--\ref{sec:execution}; App.~\ref{app:eexpl} \\
How does execution process the requests? & Fixed-request decomposition by margin; purchase-cost check & I1; recorded requests & Exploratory counterfactual & \S\ref{sec:evaluation}; App.~\ref{app:decomp}, \ref{app:headroom} \\
\midrule
\multicolumn{5}{@{}l}{\emph{Link 3 (secondary). From executed positions to conditional credit risk}} \\
When does a write-off arise? & Liquidation formula and thresholds & Any book; fixed price & Analytic & \S\ref{sec:formula} \\
Did losses occur in the markets? & Realized bad debt is 0 in all 48 markets & Both market experiments & Pre-specified descriptive; recorded in the earlier experiment & \S\ref{sec:realized} \\
Conditional losses on the main experiment's books & Fixed-price benchmark; bonus part; same-LTV counterfactual & I1; end-of-run books & Benchmark and C$-$H gap pre-specified descriptive, untested; bonus part exploratory; same-LTV analytic, fixed price only & \S\ref{sec:ebenchmark}; App.~\ref{app:eexpl} \\
Level or dispersion of leverage? & Level, threshold and dispersion metrics; reallocations & I1, I0; end books & Exploratory; reference-set positions & \S\ref{sec:ebenchmark}; App.~\ref{app:alloc} \\
Do exposure and margins vary with order? & Six sharp-null tests; LTV contrasts; post-block behavior & I1; markets, states & Planned exploratory; LTV and post-block comparisons descriptive & App.~\ref{app:planned} \\
How do liquidation rules matter? & 19-rule grid $\times$ 10, 20, 30\% shocks & I0; end-of-run books & Deterministic counterfactual; grid frozen after market outcomes, before stress outputs & \S\ref{sec:rules}; App.~\ref{app:stress} \\
How does collateral valuation matter? & Fixed-request replay, reference-valued cap & I0; recorded requests & Deterministic counterfactual (descriptive, mechanical) & App.~\ref{app:replay} \\
The same rules on the main experiment's books & Rule grid (headline rules in Table~\ref{tab:losses}; all 19 rules) and replay & I1; end-of-run books & Exploratory deterministic counterfactual & \S\ref{sec:ebenchmark}; App.~\ref{app:eexpl} \\
\midrule
\multicolumn{5}{@{}l}{\emph{Outside the main chain. Larger models}} \\
Larger models in frozen states & Larger-model frozen-state test & I0 vs.\ I1; three models; no market & Pre-specified, model by model & \S\ref{sec:scope}; App.~\ref{app:larger} \\
Larger models in news markets & News-market study, two campaigns & I1 with news; 48 markets & Pre-specified test per campaign; splits of $T$ exploratory & \S\ref{sec:scope}; App.~\ref{app:news} \\
\bottomrule
\end{tabularx}

{\scriptsize\setlength{\leftskip}{0pt}\setlength{\rightskip}{0pt}\setlength{\parfillskip}{0pt plus 1fil}\noindent\textit{Note.} Pre-specified means locked locally before the study's first model call (Section~\ref{sec:inference}). Exploratory means chosen after the results were known. A deterministic counterfactual re-executes recorded books or requests under a changed rule. The planned analyses had a plan locked in advance (Appendix~\ref{app:planned}). None of these made new model calls.\par}

\end{table}

\paragraph{Plan and timeline.} The experimental protocol, seed allocations and analysis scripts were locally hash-locked before the market experiment ran. The plan was locked after the results of the earlier experiment and the frozen-state interface test were known. The earlier experiment's post hoc directional contrast informed the test. The frozen-state interface test had met its pre-specified criterion for a revised-interface market experiment. The plan fixed one exact two-sided test of the global sharp null at $\alpha = 0.05$. It named the statistic $T$ and the estimand $\theta$, predicted $T > 0$ and set the same reporting rule for a negative $T$. It fixed the sample at 12 matched blocks in two stages of six, with 24 markets of 15 agents over 15 rounds (5,400 decisions). It allowed no extra calls or replacement markets. It also fixed $Y$ (equation~\eqref{eq:y}), invalid-output handling and technical stopping rules. It also named the descriptive outputs in advance. These were market endpoints, paired differences, group means, arm means and standard deviations. They also included final books, liquidations, realized bad debt and the request pipeline. Participation by family and mandate, one fixed-price stress benchmark and the historical comparison completed the list. Only $T$ has a pre-specified test. The only amendment came before the first call and changed audit logging only. There was no live amendment, retry or replacement.

\paragraph{Live run, markets and books.} The plan prohibited interim analysis, and the stage boundary was an operational check. Analysis and historical comparison began after all 24 markets were complete. Agents, mandates, slot rotation and book distributions follow the earlier experiment (Section~\ref{sec:market}). Books, round schedules and sampling seeds are new draws from a separate namespace (Table~\ref{tab:campaigns}). Each market starts with a fresh book, pool and feedback state. Responses are not passed as context to later requests. Initial debt averages 5,733 USDC, and block-level unused reference borrowing capacity $K_0$ ranges from 3,934 to 5,411 USDC.

\paragraph{Allocation of complete action orders.} Number the orders LDH/LHD/DLH/DHL/HLD/HDL from 0 to 5 and the blocks of a stage from 0 to 5. Balanced-market slot $j$ in block $k$ receives order $(\sigma_j + k) \bmod 6$, so every slot sees every order once per stage (Section~\ref{sec:randomization}). The balance restrictions of Appendix~\ref{app:expb} do not balance family $\times$ mandate $\times$ order cells exactly. Six fair draws per stage set the within-block run order, and the common market ran first in 7 of 12 blocks. All draws used operating-system entropy ($\tau$ through Fisher--Yates swaps). The saved allocation includes the draw indices and was never redrawn. In the realized allocation, C and H family $\times$ order and mandate $\times$ order counts match within each stage. Any family $\times$ mandate $\times$ order cell differs by at most one agent-market per stage (exploratory). The test conditions on $\sigma$ and all other inputs, including run order.

\paragraph{Reference set.} The enumeration counts ties and the observed plan among the 8,100 plans of Section~\ref{sec:estimand}. Only the observed plan and its mirror are as extreme. The enumeration reassigns complete orders within stages. Figure~\ref{fig:refdist} shows the sharp-null reference distribution, computed after the results.

\begin{center}
\captionsetup{hypcap=false}
\includegraphics[width=0.48\textwidth]{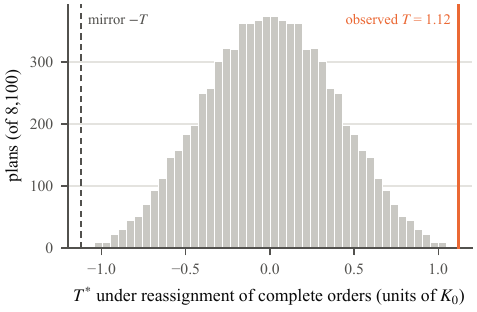}
\captionof{figure}{\textbf{Sharp-null reference distribution of $T$.} The 8,100 equally weighted plans produce a symmetric distribution with standard deviation 0.38 and a central 95 percent range of $\pm$0.74. Only the observed plan and its lever-up/delever mirror are as extreme. This is the distribution under complete-order invariance.}
\label{fig:refdist}
\end{center}

\paragraph{The 24 markets and their debt flows.} Figure~\ref{fig:paired}a shows all market endpoints. Tables of the pre-specified endpoints, paired differences and benchmark write-offs are available from the author on request. So are the exploratory borrowing and repayment flows. Hold-first group means are 0.497 (C) and 0.609 (H). No market had a liquidation or realized bad debt. The pre-specified write-off-adjusted endpoint therefore equals $Y$. Borrowing minus repayment equals the net debt change, with a maximum discrepancy of $2.7 \times 10^{-12}$ USDC across the 24 markets. Common-market borrowing averages 1.041 $K_0$ and repayment 0.335 $K_0$. Balanced-market averages are 1.084 and 0.372.

\paragraph{Decision categories and invalid outputs.} The parser follows Appendix~\ref{app:expd}. Table~\ref{tab:pipeline} reports categories and execution by family and mandate. Invalid outputs execute nothing, receive invalid-output feedback and stay in all denominators. They are never re-queried or clipped. Of 37 invalid outputs, 36 came from Llama and one from Mistral. Stage counts were 15 and 22, and no market had more than 4. All markets were retained. The inherited thresholds of 11 per market and 54 per stage served as quality diagnostics, since invalid outputs are outcomes of the interface treatment.

\paragraph{Record.} Each of the 5,400 scheduled decisions received exactly one model call and one complete response. Fixed-request replay reproduced every execution, end-of-round pool price and final borrower position under the pool-valued cap (Appendix~\ref{app:eexpl}).

\subsection{Exploratory analyses of the main experiment}\label{app:eexpl}

These analyses were chosen after the main experiment's results were known. They split debt changes into borrowing and repayment and separate debt-reducing from zero-debt collateral sales (Table~\ref{tab:pipeline}). They also cover the family split of $T$ and the fixed-request decomposition (Table~\ref{tab:decomposition}; Appendix~\ref{app:decomp}). For the loan books, they cover the benchmark's bonus component, the same-LTV counterfactual under a pool-price trigger and the full 19-rule stress grid at 10, 20 and 30 percent shocks (Table~\ref{tab:losses}). Table~\ref{tab:dexplore} gives the reason for each analysis and its headline values. Cell-level results and scripts are available from the author on request. Appendix~\ref{app:expe} lists the pre-specified test and descriptives, and Appendix~\ref{app:audit} reports the robustness and determinism checks.

\gdef\tabDexplorePositions{%
\textit{Panel A. Valid requests by the listed position of the action (percent of decisions; both arms unless stated)}\\[2pt]
\begin{tabular}{llrrrrrrrrr}
\toprule
 & & \multicolumn{4}{c}{Position of the action} & \multicolumn{5}{c}{First minus other positions (points)} \\
\cmidrule(lr){3-6}\cmidrule(lr){7-11}
Request & Family & First & Second & Third & Not first & F$-$NF & F$-$L & F$-$NF, C & F$-$NF, H & F$-$NF, std. \\
\midrule
Lever-up & All & 21.9 & 13.8 & 13.6 & 13.7 & 8.2 & 8.3 & 7.1 & 9.3 & 8.4 \\
 & Llama & 42.0 & 41.0 & 39.0 & 40.0 & 2.0 & 3.0 & 1.5 & 2.5 & 5.2 \\
 & Mistral & 23.7 & 0.3 & 1.7 & 1.0 & 22.7 & 22.0 & 19.8 & 25.5 & 20.0 \\
 & Qwen & 0.0 & 0.0 & 0.0 & 0.0 & 0.0 & 0.0 & 0.0 & 0.0 & 0.0 \\
\addlinespace
Collateral sale & All & 7.5 & 3.3 & 2.3 & 2.8 & 4.7 & 5.2 & 4.9 & 4.5 & 5.3 \\
 & Llama & 22.5 & 9.8 & 6.8 & 8.3 & 14.2 & 15.7 & 14.8 & 13.5 & 15.8 \\
\addlinespace
No trade (position of hold) & All & 84.7 & 78.4 & 72.6 & 75.5 & 9.2 & 12.1 & 10.1 & 8.2 & -- \\
 & Llama & 55.8 & 43.8 & 33.2 & 38.5 & 17.3 & 22.7 & 21.5 & 13.2 & -- \\
 & Mistral & 98.2 & 91.5 & 84.5 & 88.0 & 10.2 & 13.7 & 8.8 & 11.5 & -- \\
 & Qwen & 100.0 & 100.0 & 100.0 & 100.0 & 0.0 & 0.0 & 0.0 & 0.0 & -- \\
\bottomrule
\end{tabular}%
}

\gdef\tabDexploreCap{%
\textit{Panel B. Valid lever-up requests and the borrowing cap, by the listed position of lever-up}\\[2pt]
\begin{tabular}{lrrrrrrrr}
\toprule
 & Labeled & Valid & Executed & & Executed & \multicolumn{2}{c}{Tokens} & Fill \\
\cmidrule(lr){7-8}
 & lever-up & requests & nothing & (\%) & in part & requested & executed & (\%) \\
\midrule
Lever-up listed first & 435 & 394 & 147 & 37.3 & 25 & 883 & 499 & 56.5 \\
\quad common markets & 207 & 187 & 73 & 39.0 & 9 & 449 & 250 & 55.7 \\
\quad balanced markets & 228 & 207 & 74 & 35.7 & 16 & 434 & 249 & 57.4 \\
\quad Llama & 292 & 252 & 66 & 26.2 & 17 & 512 & 364 & 71.1 \\
\quad Mistral & 143 & 142 & 81 & 57.0 & 8 & 371 & 135 & 36.4 \\
Lever-up not listed first & 520 & 492 & 103 & 20.9 & 23 & 881 & 660 & 74.9 \\
\quad Llama & 508 & 480 & 103 & 21.5 & 23 & 848 & 627 & 73.9 \\
\quad Mistral & 12 & 12 & 0 & 0.0 & 0 & 33 & 33 & 100.0 \\
All decisions (pre-specified)$^{\ddagger}$ &  & 886 & 250 & 28.2 & 48 & 1,764 & 1,159 & 65.7 \\
\addlinespace
\multicolumn{9}{l}{Lever-up tokens per decision, listed first vs not: requested 0.49 vs 0.24; executed 0.28 vs 0.18.} \\
\multicolumn{9}{l}{Ratio of the executed to the requested token contrast: 0.38 (common markets 0.41, balanced markets 0.36).} \\
\bottomrule
\end{tabular}%
}

\gdef\tabDexploreGroups{%
\textit{Panel C. Common markets by the first action of their complete order}\\[2pt]
\begin{tabular}{>{\raggedright\arraybackslash}p{7.6cm}rrrr}
\toprule
 & Lever-up first & Hold first & Delever first & All 12 \\
\midrule
Complete orders & LHD, LDH & HLD, HDL & DLH, DHL &  \\
Mean $Y$ (pre-specified)$^{\ddagger}$ & 1.225 & 0.497 & 0.397 & 0.706 \\
SD of $Y$ within the group (pooled for all 12) & 0.48 & 0.05 & 0.16 & 0.29 \\
Balanced markets above the group's lowest common market & 8 & 10 & 12 &  \\
Balanced markets below the group's highest common market & 12 & 3 & 3 &  \\
Valid lever-up requests (\% of decisions) & 20.8 & 10.3 & 17.0 &  \\
Valid collateral-sale requests (\% of decisions) & 2.1 & 3.3 & 7.7 &  \\
\addlinespace
Between-group share of the common-market sum of squares (\%) &  &  &  & 68.1 \\
SD of $Y$: common / balanced (ratio; pre-specified)$^{\ddagger}$ &  &  &  & 0.47 / 0.23 (2.05) \\
SD ratio, stage 1 / stage 2 &  &  &  & 1.94 / 2.16 \\
\bottomrule
\end{tabular}%
}

\gdef\tabDexploreFamily{%
\textit{Panel D. Family accounting split, group means of the within-block difference C $-$ H per market}\\[2pt]
\begin{tabular}{llrrrrr}
\toprule
 & & Lever-up & Delever & Hold & Lever-up first & Share \\
Quantity & Family & first & first & first & minus delever first & (\%) \\
\midrule
Net debt change / $K_0$ & Llama & 0.444 & $-$0.279 & $-$0.084 & 0.723 & 64.5 \\
 & Mistral & 0.164 & $-$0.234 & $-$0.028 & 0.398 & 35.5 \\
 & Qwen & 0.000 & 0.000 & 0.000 & 0.000 &  \\
 & All (= $T$)$^{\ddagger}$ & 0.608 & $-$0.513 & $-$0.112 & 1.121 &  \\
\addlinespace
Valid lever-up requests & Llama & 5.00 & 4.00 & $-$9.00 & 1.00 & 5.6 \\
 & Mistral & 7.50 & $-$9.25 & $-$3.25 & 16.75 & 94.4 \\
 & All & 12.50 & $-$5.25 & $-$12.25 & 17.75 &  \\
\addlinespace
Valid collateral-sale requests & Llama (all) & $-$6.25 & 8.75 & $-$2.25 & $-$15.00 &  \\
\addlinespace
Executed lever-up tokens & Llama & 10.25 & $-$6.50 & $-$6.00 & 16.75 & 48.9 \\
 & Mistral & 7.25 & $-$10.25 & $-$1.50 & 17.50 & 51.1 \\
 & All & 17.50 & $-$16.75 & $-$7.50 & 34.25 &  \\
\addlinespace
Executed collateral-sale tokens & Llama (all) & $-$11.00 & 11.25 & $-$2.00 & $-$22.25 &  \\
\addlinespace
Mean quantity per valid lever-up request, C / H & Llama & 2.19 / 1.90 & 1.52 / 1.74 & 2.23 / 1.77 & & \\
\bottomrule
\end{tabular}%
}

\gdef\tabDexploreFamilyTable{%
\begin{tabular}{llrrrrr}
\toprule
 & & Lever-up & Delever & Hold & Lever-up first & Share \\
Quantity & Family & first & first & first & minus delever first & (\%) \\
\midrule
Net debt change / $K_0$ & Llama & 0.444 & $-$0.279 & $-$0.084 & 0.723 & 64.5 \\
 & Mistral & 0.164 & $-$0.234 & $-$0.028 & 0.398 & 35.5 \\
 & Qwen & 0.000 & 0.000 & 0.000 & 0.000 &  \\
 & All (= $T$)$^{\ddagger}$ & 0.608 & $-$0.513 & $-$0.112 & 1.121 &  \\
\addlinespace
Valid lever-up requests & Llama & 5.00 & 4.00 & $-$9.00 & 1.00 & 5.6 \\
 & Mistral & 7.50 & $-$9.25 & $-$3.25 & 16.75 & 94.4 \\
 & All & 12.50 & $-$5.25 & $-$12.25 & 17.75 &  \\
\addlinespace
Valid collateral-sale requests & Llama (all) & $-$6.25 & 8.75 & $-$2.25 & $-$15.00 &  \\
\addlinespace
Executed lever-up tokens & Llama & 10.25 & $-$6.50 & $-$6.00 & 16.75 & 48.9 \\
 & Mistral & 7.25 & $-$10.25 & $-$1.50 & 17.50 & 51.1 \\
 & All & 17.50 & $-$16.75 & $-$7.50 & 34.25 &  \\
\addlinespace
Executed collateral-sale tokens & Llama (all) & $-$11.00 & 11.25 & $-$2.00 & $-$22.25 &  \\
\addlinespace
Mean quantity per valid lever-up request, C / H & Llama & 2.19 / 1.90 & 1.52 / 1.74 & 2.23 / 1.77 & & \\
\bottomrule
\end{tabular}%
}

\gdef\tabDexploreByFirst{%
\textit{Panel A. End-of-run positions and creditor exposure by the first action of the block's common order (USDC per market unless stated)}\\[2pt]
\begin{tabular}{>{\raggedright\arraybackslash}p{6.0cm}rrrrrrrr}
\toprule
 & \multicolumn{4}{c}{Common markets} & \multicolumn{4}{c}{Balanced markets of the same blocks} \\
\cmidrule(lr){2-5}\cmidrule(lr){6-9}
 & L first & D first & H first & All & L first & D first & H first & All \\
\midrule
Borrowers above 0.648 (write-off zone at 30\%) & 19 & 9 & 10 & 38 & 12 & 16 & 12 & 40 \\
Borrowers above 0.70 & 16 & 2 & 4 & 22 & 4 & 8 & 5 & 17 \\
Largest borrower reference LTV & 0.732 & 0.704 & 0.714 & 0.732 & 0.716 & 0.718 & 0.719 & 0.719 \\
Mean aggregate reference LTV$^{\ddagger}$ & 0.559 & 0.470 & 0.470 & 0.500 & 0.496 & 0.525 & 0.486 & 0.502 \\
Mean peak pool-price run-up (\%) & 5.1 & 2.4 & 2.4 & 3.3 & 2.8 & 3.7 & 3.4 & 3.3 \\
Mean end-of-run debt$^{\ddagger}$ & 11,435 & 7,218 & 8,330 & 8,994 & 8,687 & 9,524 & 8,917 & 9,043 \\
\addlinespace
Scenario write-off $B(70)$ (no feedback, 30\%)$^{\ddagger}$ & 630 & 151 & 189 & 323 & 225 & 352 & 269 & 282 \\
\quad of which pre-liquidation deficiency $U_0(70)$$^{\ddagger}$ & 158 & 3 & 11 & 57 & 20 & 38 & 22 & 27 \\
\quad bonus component $B(70) - U_0(70)$ &  &  &  & 266 &  &  &  & 255 \\
\quad write-offs on loans covered at $F$ before liquidation (borrowers) &  &  &  & 80 (16) &  &  &  & 117 (23) \\
\quad as \% of pre-shock debt (pooled) & 5.5 & 2.1 & 2.3 & 3.6 & 2.6 & 3.7 & 3.0 & 3.1 \\
External trigger, pool sale, 30\% & 350 & 132 & 163 & 215 & 151 & 201 & 181 & 178 \\
Pool trigger, pool sale, 20\% & 399 & 0 & 2 & 134 & 13 & 49 & 21 & 28 \\
Pool trigger, pool sale, 30\% & 1,210 & 229 & 318 & 586 & 418 & 636 & 466 & 507 \\
\bottomrule
\end{tabular}%
}

\gdef\tabDexploreHeadline{%
\textit{Panel B. Headline rules (profit-seeking liquidators), 24 main experiment's books (USDC per market)}\\[2pt]
\begin{tabular}{ll*{12}{>{\raggedleft\arraybackslash}p{0.74cm}}}
\toprule
 & & \multicolumn{2}{c}{Creditor} & \multicolumn{2}{c}{of which} & \multicolumn{4}{c}{Exposure / debt (\%)} & \multicolumn{2}{c}{Mean of} & \multicolumn{2}{c}{Liquidation} \\
 & & \multicolumn{2}{c}{exposure $U(F)$} & \multicolumn{2}{c}{remaining def.} & \multicolumn{2}{c}{pooled} & \multicolumn{2}{c}{by market} & \multicolumn{2}{c}{worst 3} & \multicolumn{2}{c}{events} \\
\cmidrule(lr){3-4}\cmidrule(lr){5-6}\cmidrule(lr){7-8}\cmidrule(lr){9-10}\cmidrule(lr){11-12}\cmidrule(lr){13-14}
Rule & Shock & C & H & C & H & C & H & C & H & C & H & C & H \\
\midrule
\textit{No liquidation: $U_0$} & 10\% & 0 & 0 & -- & -- & -- & -- & -- & -- & -- & -- & -- & -- \\
 & 20\% & 0 & 0 & -- & -- & -- & -- & -- & -- & -- & -- & -- & -- \\
 & 30\% & 57 & 27 & -- & -- & -- & -- & -- & -- & -- & -- & -- & -- \\
\addlinespace
No feedback & 10\% & 0 & 0 & 0 & 0 & 0.0 & 0.0 & 0.0 & 0.0 & 0 & 0 & 0.8 & 0.0 \\
 & 20\% & 0 & 0 & 0 & 0 & 0.0 & 0.0 & 0.0 & 0.0 & 0 & 0 & 7.1 & 6.4 \\
 & 30\% & 323 & 282 & 0 & 0 & 3.6 & 3.1 & 3.2 & 3.0 & 791 & 410 & 15.5 & 17.4 \\
\addlinespace
External trigger, pool sale & 10\% & 0 & 0 & 0 & 0 & 0.0 & 0.0 & 0.0 & 0.0 & 0 & 0 & 0.8 & 0.0 \\
 & 20\% & 0 & 0 & 0 & 0 & 0.0 & 0.0 & 0.0 & 0.0 & 0 & 0 & 4.2 & 5.3 \\
 & 30\% & 215 & 178 & 174 & 157 & 2.4 & 2.0 & 2.2 & 1.9 & 417 & 254 & 7.5 & 7.2 \\
\addlinespace
Pool trigger, pool sale & 10\% & 0 & 0 & 0 & 0 & 0.0 & 0.0 & 0.0 & 0.0 & 0 & 0 & 2.8 & 0.0 \\
 & 20\% & 134 & 28 & 0 & 0 & 1.5 & 0.3 & 1.0 & 0.3 & 532 & 79 & 12.3 & 18.1 \\
 & 30\% & 586 & 507 & 0 & 0 & 6.5 & 5.6 & 5.6 & 5.4 & 1,547 & 747 & 18.2 & 19.4 \\
\addlinespace
\quad pool replenished & 10\% & 0 & 0 & 0 & 0 & 0.0 & 0.0 & 0.0 & 0.0 & 0 & 0 & 0.8 & 0.0 \\
 & 20\% & 20 & 0 & 0 & 0 & 0.2 & 0.0 & 0.2 & 0.0 & 81 & 0 & 10.2 & 7.7 \\
 & 30\% & 408 & 352 & 0 & 0 & 4.5 & 3.9 & 3.9 & 3.8 & 1,059 & 521 & 15.8 & 17.7 \\
\bottomrule
\end{tabular}%
}

\gdef\tabDexploreOther{%
\textit{Panel C. Other rules of the frozen grid at 20 and 30 percent shocks (USDC per market)}\\[2pt]
\begin{tabular}{lrrrrllrrr}
\toprule
 & \multicolumn{2}{c}{Mean, 20\%} & \multicolumn{2}{c}{Mean, 30\%} & \multicolumn{2}{c}{Write-off / deficiency, 30\%} & \multicolumn{2}{c}{Worst 3 of 12, 30\%} & Ratio \\
\cmidrule(lr){2-3}\cmidrule(lr){4-5}\cmidrule(lr){6-7}\cmidrule(lr){8-9}
Rule & C & H & C & H & C & H & C & H & C/H \\
\midrule
Pool trigger, pool sale, unconstrained & 134 & 28 & 586 & 507 & 586/0 & 507/0 & 1,547 & 747 & 2.07 \\
External trigger, pool sale, unconstrained & 0 & 0 & 323 & 282 & 323/0 & 282/0 & 791 & 410 & 1.93 \\
No feedback, unconstrained & 0 & 0 & 323 & 282 & 323/0 & 282/0 & 791 & 410 & 1.93 \\
Pool trigger, capital-limited & 3 & 13 & 402 & 456 & 199/203 & 250/206 & 812 & 620 & 1.31 \\
External trigger, capital-limited & 0 & 0 & 215 & 178 & 41/174 & 21/157 & 417 & 254 & 1.64 \\
No feedback, capital-limited & 0 & 0 & 264 & 271 & 126/138 & 187/84 & 553 & 378 & 1.46 \\
Blend oracle 0.25, unconstrained & 5 & 0 & 384 & 334 & 384/0 & 334/0 & 970 & 494 & 1.96 \\
Blend oracle 0.50, unconstrained & 41 & 0 & 450 & 390 & 450/0 & 390/0 & 1,156 & 578 & 2.00 \\
Blend oracle 0.75, unconstrained & 85 & 6 & 516 & 447 & 516/0 & 447/0 & 1,345 & 659 & 2.04 \\
Blend oracle 0.25, profitable & 0 & 0 & 269 & 258 & 130/139 & 43/215 & 513 & 349 & 1.47 \\
Blend oracle 0.50, profitable & 1 & 0 & 350 & 385 & 161/189 & 362/24 & 757 & 561 & 1.35 \\
Blend oracle 0.75, profitable & 85 & 6 & 516 & 447 & 516/0 & 447/0 & 1,345 & 659 & 2.04 \\
External trigger, full close below HF 0.95 & 0 & 0 & 224 & 202 & 189/36 & 196/6 & 433 & 307 & 1.41 \\
Pool trigger, full close below HF 0.95 & 121 & 26 & 543 & 467 & 543/0 & 467/0 & 1,466 & 681 & 2.15 \\
Pool trigger, replenished, unconstrained & 20 & 0 & 408 & 352 & 408/0 & 352/0 & 1,059 & 521 & 2.03 \\
\addlinespace
\multicolumn{10}{l}{\textit{Shock relative to the end-of-run pool price, $F = p_{\text{end}}(1-s)$}} \\
\quad No feedback & 0 & 0 & 299 & 261 & 299/0 & 261/0 & 718 & 359 &  \\
\quad External trigger, pool sale & 0 & 0 & 190 & 160 & 32/158 & 20/140 & 340 & 229 &  \\
\quad Pool trigger, pool sale & 115 & 16 & 556 & 484 & 556/0 & 484/0 & 1,469 & 705 &  \\
\bottomrule
\end{tabular}%
}

\gdef\tabDexploreBenchmarks{%
\textit{Panel A. Benchmark books, rule contrasts, tails and checks (USDC per market unless stated)}\\[2pt]
\begin{tabular}{>{\raggedright\arraybackslash}p{8.2cm}ll}
\toprule
 & Common (C) & Balanced (H) \\
\midrule
\multicolumn{3}{l}{\textit{Benchmark books}} \\
Initial books, all rules and shocks (684 scenarios): largest exposure & \multicolumn{2}{c}{0} \\
Every borrower at reference LTV 0.55 / 0.65 / 0.70, no feedback, 30\% & \multicolumn{2}{c}{0 / 28 / 770} \\
Debt re-spread at each market's aggregate reference LTV: largest aggregate LTV$^{\ddagger}$ & 0.620 & 0.538 \\
\quad markets with exposure at 30\%, no feedback or external trigger & 0 & 0 \\
\quad markets with exposure at 30\%, pool trigger (blocks) & 2 (2, 11) & 0 \\
\quad exposure in those markets, pool trigger, 30\% & 802; 1,323 & -- \\
\quad rules (of 19) with no exposure in any of the 24 books, 10\% / 20\% / 30\% & \multicolumn{2}{c}{19 / 19 / 10} \\
\addlinespace
\multicolumn{3}{l}{\textit{Rule contrasts and liquidations (profit-seeking liquidators; mean per market)}} \\
Pool trigger minus external trigger, 20\% / 30\% & 134 / 371 & 28 / 329 \\
External trigger (pool sale) minus no feedback, 30\% & $-$108 & $-$104 \\
External trigger, 30\%: write-offs & 41 & 21 \\
\quad accounts still liquidatable at $F$ & 2.9 & 3.5 \\
\quad liquidation transfer from borrowers & 221 & 230 \\
\addlinespace
\multicolumn{3}{l}{\textit{Tails at 30\% (worst three of 12 markets)}} \\
Blocks of the worst three: no feedback / external / pool trigger & 11,2,5 / 11,2,5 / 2,11,5 & 7,4,2 / 4,7,1 / 7,2,1 \\
Ratio of worst-three means C/H, range over the 19 rules & \multicolumn{2}{c}{1.31--2.15} \\
\quad leaving out one block: no feedback / external / pool trigger & \multicolumn{2}{c}{1.39--2.16 / 1.28--1.85 / 1.49--2.34} \\
Without the four lever-up-first blocks: worst three, no feedback & 232 & 404 \\
\quad mean, no feedback & 170 & 311 \\
\addlinespace
\multicolumn{3}{l}{\textit{Checks}} \\
No-feedback identity: largest $|U(F) - B(F)|$ over 720 scenarios (USDC) & \multicolumn{2}{c}{$5.7\times 10^{-13}$} \\
Accounting identity: largest residual over 5,904 scenarios (USDC) & \multicolumn{2}{c}{$1.6\times 10^{-10}$} \\
Pool-trigger rule: profit-seeking minus unconstrained liquidators, all books and shocks & \multicolumn{2}{c}{0} \\
\bottomrule
\end{tabular}%
}

\gdef\tabDexploreReplay{%
\textit{Panel B. Spot-valued versus reference-valued borrowing cap in a fixed-request replay of the main experiment's records}\\[2pt]
\begin{tabular}{>{\raggedright\arraybackslash}p{6.6cm}ll}
\toprule
 & Spot-valued cap & Reference-valued cap \\
 & (recorded; replay exact) & (same requests, replayed) \\
\midrule
$Y$: mean, C / H & 0.706 / 0.712 & 0.633 / 0.652 \\
$Y$: SD, C / H (ratio) & 0.47 / 0.23 (2.05) & 0.39 / 0.21 (1.83) \\
Lever-up-first minus delever-first contrast in C $-$ H ($T$ on replayed endpoints) & 1.121 & 0.936 \\
$V_{20}$: SD ratio C / H & 2.38 & 2.23 \\
Largest borrower LTV at 100, C / H & 0.732 / 0.719 & 0.698 / 0.700 \\
Borrowers above 0.70, C / H & 22 / 17 & 0 / 0 \\
Borrowers above 0.648, C / H & 38 / 40 & 37 / 40 \\
Peak pool-price run-up, lever-up-first C markets (\%, range) & 3.0--6.2 & 2.8--5.8 \\
Peak pool-price run-up, largest balanced market (\%) & 4.3 & 4.1 \\
Valid lever-up requests, C / H & 433 / 453 & 433 / 453 \\
\quad executing nothing, C / H & 130 / 120 & 140 / 134 \\
Executed lever-up tokens, C / H & 566 / 593 & 528 / 559 \\
End debt, mean per market, C / H & 8,994 / 9,043 & 8,662 / 8,766 \\
\addlinespace
\multicolumn{3}{l}{\textit{Creditor exposure, C / H (USDC per market; \% of pre-shock debt, pooled)}} \\
\quad No feedback, 30\%$^{\ddagger}$: mean & 323 / 282 (3.6 / 3.1) & 205 / 202 (2.4 / 2.3) \\
\quad\quad per unit of $K_0$ & 0.0705 / 0.0607 & 0.0445 / 0.0436 \\
\quad\quad lever-up-first common markets & 630 & 366 \\
\quad\quad worst 3 of 12 & 791 / 410 & 451 / 315 \\
\quad External trigger, pool sale, 30\%: mean & 215 / 178 (2.4 / 2.0) & 121 / 112 (1.4 / 1.3) \\
\quad\quad worst 3 of 12 & 417 / 254 & 181 / 165 \\
\quad\quad write-off / deficiency, C and H & 41/174 and 21/157 & 35/86 and 14/98 \\
\quad Pool trigger, pool sale, 30\%: mean & 586 / 507 (6.5 / 5.6) & 432 / 409 (5.0 / 4.7) \\
\quad\quad worst 3 of 12 & 1,547 / 747 & 1,093 / 630 \\
\quad Pool trigger, pool sale, 20\%: mean & 134 / 28 & 47 / 4 \\
\bottomrule
\end{tabular}%
}

\gdef\tabDexploreSummary{%
\begin{longtable}{@{}>{\raggedright\arraybackslash}p{\dimexpr0.25\textwidth-\tabcolsep\relax}>{\raggedright\arraybackslash}p{\dimexpr0.75\textwidth-\tabcolsep\relax}@{}}
\caption{\textbf{Summary of the exploratory analyses of the main experiment's records.} One row per analysis. C denotes common markets and H balanced markets. Exposure is creditor exposure $U(F)$ \eqref{eq:exposure} at $F = 100(1-s)$ with profit-seeking liquidators, under the 19 rules frozen for the earlier experiment's books (Appendix~\ref{app:stress}) and applied unchanged. Entries marked $\ddagger$ repeat pre-specified outputs (Tables~\ref{tab:emain} and~\ref{tab:losses}) or, for the largest aggregate reference LTV, the maximum of a pre-specified per-market output. Full tables, cell-level values and notes are available from the author on request. Differences are computed before rounding.}\label{tab:dexplore}\\
\toprule
Analysis (why reported) & Main-experiment records (C / H; USDC per market unless stated) \\
\midrule
\endfirsthead
\multicolumn{2}{@{}l}{\textit{Table~\thetable\ (continued)}}\\[2pt]
\toprule
Analysis (why reported) & Main-experiment records (C / H; USDC per market unless stated) \\
\midrule
\endhead
\bottomrule
\endlastfoot
Requests by listed position (size of the position effect in the markets; Section~\ref{sec:requests}) & Valid lever-up request in 21.9\% of decisions with lever-up listed first and 13.7\% otherwise (8.2 points; 7.1 in C, where the position is shared and the difference includes market feedback, 9.3 in H). Mistral 23.7 vs.\ 1.0\%; Llama 42.0 / 41.0 / 39.0\% by position; Qwen none. Collateral-sale requests (Llama only) 22.5 / 9.8 / 6.8\% by position of delever; no trade 84.7 / 78.4 / 72.6\% by position of hold \\
\addlinespace
The cap and first-listed requests (where the cap binds; Section~\ref{sec:execution}) & Executed nothing: 147 of 394 first-listed valid lever-up requests (37.3\%; Mistral 57.0\%, Llama 26.2\%) vs.\ 103 of 492 otherwise (20.9\%); tokens bought 56.5 vs.\ 74.9\%. Ratio of the executed to the requested lever-up token contrast 0.38 (C 0.41, H 0.36) \\
\addlinespace
Dispersion among common markets (how much of the SD ratio 2.05$^{\ddagger}$ lies between first-action groups; Section~\ref{sec:reading}) & Between-group share of the common-market sum of squares of $Y$ 68.1\%. Within-group SD: lever-up first 0.48, delever first 0.16, hold first 0.05; pooled 0.29 vs.\ 0.23$^{\ddagger}$ across H. 8 of 12 H markets lie above the lowest lever-up-first C market. SD ratio by stage 1.94 / 2.16 \\
\addlinespace
Family accounting split of $T$ (participation is concentrated in two families; Section~\ref{sec:participation}) & Net debt change$/K_0$: Llama 0.723, Mistral 0.398, Qwen 0.000 (sum $T = 1.121$$^{\ddagger}$). Valid lever-up requests 17.75 per market (Mistral 16.75, Llama 1.00); Llama valid collateral-sale requests $-$15.00; executed lever-up tokens Llama 16.75, Mistral 17.50. The full split by first-action group is available on request \\
\addlinespace
End-of-run positions (the borrower-level allocation behind the fixed-price benchmark; Section~\ref{sec:ebenchmark}) & Reference LTV above 0.648: 38 borrowers in C (19 in lever-up-first C) vs.\ 40 in H; above 0.70: 22 (16) vs.\ 17; largest 0.732 / 0.719. Benchmark write-offs 323$^{\ddagger}$ / 282$^{\ddagger}$ (lever-up-first C 630), of which pre-liquidation deficiency 57$^{\ddagger}$ / 27$^{\ddagger}$ and bonus component 266 / 255; write-offs on loans covered at $F$ before liquidation 80 (16 borrowers) / 117 (23) \\
\addlinespace
Headline rules at 30\% (how the frozen rules of Section~\ref{sec:rules} act on these books) & No feedback 323$^{\ddagger}$ / 282$^{\ddagger}$ (3.6 / 3.1\% of pre-shock debt pooled, 3.2 / 3.0\% by market); external trigger 215 / 178 (write-offs 41 / 21, remaining deficiency 174 / 157; 2.9 / 3.5 accounts still liquidatable; liquidation transfer 221 / 230); pool trigger 586 / 507; pool replenished 408 / 352 \\
\addlinespace
Moderate shocks (whether 10 and 20\% shocks reach creditors) & 10\%: no exposure under any rule (pool trigger: 2.8 / 0.0 liquidations per market). 20\%: none with no feedback or an external trigger; pool trigger 134 / 28 (lever-up-first C 399), 20 / 0 with the pool replenished \\
\addlinespace
Benchmark and re-spread books (whether exposure comes from executed positions; scope of the same-LTV counterfactual, Section~\ref{sec:ebenchmark}) & Initial books: no exposure in 684 scenarios. Every borrower at reference LTV 0.55 / 0.65 / 0.70, no feedback, 30\%: 0 / 28 / 770. Debt re-spread at each market's aggregate reference LTV (at most 0.620$^{\ddagger}$ in C, 0.538$^{\ddagger}$ in H): no exposure at 30\% with no feedback or an external trigger; with a pool trigger, exposure stays in 2 lever-up-first C markets (802 and 1,323); rules (of 19) leaving no exposure in any re-spread book: 19 / 19 / 10 at 10 / 20 / 30\% \\
\addlinespace
Tails and the other 15 rules (where exposure sits among 12 books per arm) & Worst three C markets lever-up first under each headline rule; worst-three means 791 / 410 (no feedback), 417 / 254 (external), 1,547 / 747 (pool); ratio 1.31--2.15 over the 19 rules. Without the four lever-up-first blocks, no feedback: worst three 232 / 404, mean 170 / 311. Mean at 30\% lower in C under 3 of the 19 rules (capital-limited pool trigger and no feedback; profit-seeking blend oracle 0.50). Shock from the end-of-run pool price: levels change, not the ordering of the arms \\
\addlinespace
Fixed-request replay with a reference-valued cap (whether it removes the positions above 0.70; Section~\ref{sec:evaluation} and Appendix~\ref{app:replay}) & The spot-valued replay reproduces every recorded execution, price and book; no liquidation under either cap. Reference-valued cap: requests executing nothing 130 $\to$ 140 (C), 120 $\to$ 134 (H); above 0.70 22 / 17 $\to$ 0 / 0 (largest 0.698 / 0.700); above 0.648, 37 / 40 remain. Exposure at 30\%: no feedback 205 / 202, external 121 / 112, pool 432 / 409. Contrast on replayed endpoints 0.936 (pre-specified $T = 1.121$$^{\ddagger}$); SD ratio 1.83 (2.05$^{\ddagger}$). Agents do not react to the replayed states \\
\end{longtable}%
}

{\scriptsize
\setlength{\tabcolsep}{3pt}
\setlength{\LTcapwidth}{\textwidth}
\renewcommand{\arraystretch}{1.15}
\tabDexploreSummary
}

\section{The earlier experiment}\label{app:expb}

The earlier experiment had 12 matched blocks and 5,400 decisions under I0 (Table~\ref{tab:campaigns}). Figure~\ref{fig:paired}b summarizes its outcomes, and Appendix~\ref{app:history} compares the market experiments. Per-market and group outcomes, requests and execution, and invalid-output categories are available from the author on request. So are the pre-specified interval and constant-shift calculations and the post hoc enumeration.

\paragraph{Design, plan and record.} Families and mandates rotate across agent slots and blocks, with five of each in every market (Section~\ref{sec:market}). Within each six-block stage, one uniformly drawn one-to-one assignment gave the six complete orders to the six common markets. The balanced offsets were drawn uniformly from 165,888 vectors. Under each of these vectors, every slot sees every order once in each arm and stage. Every family and every mandate sees every order five times, and every first-listed action occurs 30 times. Three-way balance is infeasible. The arms therefore differ in the family $\times$ mandate $\times$ order cells by at most two agent-markets in Stage 1 and one in Stage 2. The run order of a block's two markets was a fair draw. The specification, allocation, inference plan and code were hashed locally before the first call.

Gate G2 capped invalid outputs at 2 percent of a stage's decisions and 5 percent of any market's. Stage 1 passed (44 invalid outputs, 1.6 percent). In Stage 2, the common market of block 9 (order DHL) produced 12 invalid outputs (5.3 percent) against a per-market limit of 11. The stop rule halted the stage after 8 of 12 markets. An amendment turned G2 from a stopping rule into a mandatory diagnostic and changed nothing else. It was locked before any further call and before any Stage 2 financial outcome was examined. The remaining four markets were run, and the pre-specified analysis was run once on all 24. First-action outcomes were pre-specified descriptives. The rankings, variance split and randomization calculation below are post hoc. Across all 5,400 calls there were no aborted calls, retries or re-run markets. Replay reproduced every decision, round and final book, and no market had a liquidation or realized bad debt.

\paragraph{Market outcomes.} Every lever-up-first common market ended above every balanced market, and every delever-first common market ended below them (Figure~\ref{fig:paired}b; Table~\ref{tab:history}). Hold-first outcomes were near zero. Paired differences average 1.35 in lever-up-first blocks and $-1.35$ in delever-first blocks. Common-market $Y$ has standard deviation 1.19 against 0.39 in balanced markets, a ratio of 3.02. First-action groups account for 95.6 percent of the common-arm sum of squares. The pooled within-group standard deviation is 0.27, below the balanced 0.39. The ratio thus describes spread across orders, not volatility under a single shared order. Under the spot-valued cap \eqref{eq:cap}, purchases raise the price used to value all collateral. Pool prices peaked 6.6 to 10.0 percent above reference in lever-up-first common markets, against at most 4.5 percent in balanced markets. The lever-up-first common markets ended with 32 borrowers above reference LTV 0.70 (maximum 0.753). Balanced markets had 30 such borrowers across ten markets (none above 0.723). Every new borrowing trade complied with the 70 percent cap, and no liquidation occurred during the runs.

\paragraph{Inference.} The pre-specified statistic is the Pitman--Morgan correlation $r = \operatorname{corr}(Y^C + Y^H, Y^C - Y^H)$ across the 12 blocks. Its sign is that of the sample variance difference, because $\operatorname{cov}(Y^C + Y^H, Y^C - Y^H) = s_C^2 - s_H^2$. Its reference distribution is the set of all $2^{12}$ within-block swaps of the C and H labels. The observed $r = 0.803$ is matched or exceeded by 11 of the 4,096 swaps (one-sided $p = 11/4{,}096 = 0.0027$). C and H roles were fixed. Only the within-block run order was randomized. The test is exact under joint exchangeability of the 12 pairs. The paired mean difference is $-0.055$ (two-sided sign-flip $p = 0.88$). The pre-specified approximate 90 percent interval for the standard-deviation ratio and the constant-shift set for the mean difference are available on request. A post hoc randomization calculation over the 8,100 first-action labelings of Section~\ref{sec:estimand} is also available. In that calculation, the observed lever-up-first minus delever-first difference in mean $Y$ is 2.65. This is the largest of the 8,100 values under the sharp null of order invariance (one-sided $p = 1/8{,}100$).

\paragraph{From requests to executed positions.} Valid lever-up requests occur in 33 percent of first-listed decisions (601 of 1,800) and 7 percent otherwise (243 of 3,600). Delever is proposed in 26 percent of first-listed decisions. Among 617 first-listed lever-up proposals, 16 are invalid. Of the 601 valid requests, the cap blocks 238 (39.6 percent) and partially executes 33. Executed tokens are 54 percent of those requested when lever-up is first and 62 percent across all valid requests. Among 467 first-listed delever proposals, 61 (13 percent) are invalid. These include 27 zero-quantity outputs that the prompt allowed but the parser rejected. Another 57 valid requests execute nothing because collateral is less than one whole token. First-listed lever-up decisions request 0.60 more tokens (0.72 against 0.12) and execute 0.28 more (0.39 against 0.11). Their descriptive contrast ratio is 0.48 (0.49 in common and 0.47 in balanced markets). Mistral has the greatest order sensitivity and the highest cap-blocking rate. All 86 invalid outputs (1.6 percent of decisions) name a valid action with a rejected quantity, and 41 of them are zero. They execute as hold and stay in all denominators. Valid-request statistics count positive quantities. Appendix~\ref{app:expd} re-parses the zeros under I1.

\section{Stress rules, identities and the fixed-request replay}\label{app:stress}

The stress grid applies deterministic counterfactuals to the 24 terminal books of the earlier experiment. Its specification was frozen after the market outcomes were known but before any stress output was computed. The no-feedback identity and thresholds hold for any book. Applying the grid to the main experiment's books is exploratory (Appendix~\ref{app:eexpl}). The full grid, pass-through and liquidation-transfer comparisons, same-LTV benchmark books and wealth decompositions are available from the author on request.

\paragraph{Rules.} The frozen stress specification defines 19 liquidation rules. All share an 80 percent liquidation threshold, an 8 percent bonus and a 0.3 percent AMM fee. All have no interest and passive borrowers and run liquidation to a fixed point (a 200-round limit, never reached). The rules differ in five ways. The trigger price is the pool price, the external price $F$ or a geometric blend $F^{1-\varphi}\,p^{\varphi}$ with pass-through $\varphi \in \{0.25, 0.5, 0.75\}$. Seized collateral is sold into the pool or outside at $F$. Liquidators are unconstrained or profit-seeking. Profit-seeking liquidators act only if the sale covers the repayment, and one variant gives them an illustrative 5,000 USDC of aggregate capital. The close factor is 50 percent, or a full close below a health factor of 0.95. Between liquidation rounds there is no arbitrage, or the pool is replenished to $F$. Each rule is applied at shocks of 10, 20 and 30 percent to the reference price, $F = 100(1-s)$. With the benchmark books, the stress outputs comprise 3,096 scenarios. The three headline rules are those of Section~\ref{sec:rules}. The specification states the pool-price rule with unconstrained liquidators. On these books its profit-seeking and unconstrained versions give identical exposure at every shock. The headline rules were also evaluated with the shock applied to the end-of-run pool price. Table~\ref{tab:losses} and the main text use the reference base.

\paragraph{The no-feedback identity.} Equation~\eqref{eq:bonus} follows from the invariant \eqref{eq:invariant} under five conditions. Liquidation is triggered and collateral seized at the price $F$. The liquidator either sells outside at $F$ or bears any loss from selling into the pool. Liquidation continues until no account can be liquidated. Borrowers are passive and pay no interest. The liquidation threshold (0.80) is below $1/1.08$, so every account with a positive invariant can be liquidated. For a fixed partial close factor $0 < k < 1$, debt falls geometrically, and the invariant decides the outcome. If $d > Fc/1.08$, the wipe-out condition $c \le 1.08\,k d/F$ is eventually met. Total recovery is then $Fc/1.08$, and the invariant is written off. If $d < Fc/1.08$, the account is or becomes healthy after finitely many partial liquidations, and the remaining loan is covered. At the boundary $d = Fc/1.08$, partial liquidation keeps the LTV at $1/1.08 > 0.80$ and never reaches the healthy region in finitely many steps. Debt and collateral still converge to zero, and the limiting write-off is zero, as under a full close. No remaining account has a collateral deficiency.

On all 720 no-feedback scenarios, the largest absolute discrepancy from the engine is $1.5\times 10^{-12}$ USDC. In each of the 3,096 scenarios, the wealth changes of all parties sum to the shock's revaluation of collateral and pool inventory (largest residual $1.8 \times 10^{-10}$ USDC). The parties are borrowers, creditors, liquidators, arbitrageurs and the pool. Re-execution from the saved records reproduces every recorded scenario to within $5\times 10^{-13}$ USDC. The thresholds \eqref{eq:thresholds} for liquidation, write-off and deficiency are 0.720, 0.833 and 0.900 at a 10 percent shock. They are 0.640, 0.741 and 0.800 at 20 percent, and 0.560, 0.648 and 0.700 at 30 percent. The initial books generate no creditor exposure under any rule at any shock.

\paragraph{Headline rules on the earlier experiment's books.} Three rules apply profit-seeking liquidation to the terminal books (Table~\ref{tab:lossesorig}; Figure~\ref{fig:stress}c--e). The \emph{no-feedback} rule triggers and settles at $F$. The \emph{external-price trigger} settles in the pool, and so does the \emph{pool-price trigger}. At a 30 percent shock, no-feedback write-offs average 471 USDC per common market and 371 per balanced market. Reallocating each market's debt in proportion to collateral equalizes LTV (Section~\ref{sec:ebenchmark}). Under the no-feedback rule, this removes all balanced-market exposure and leaves one common market exposed, at aggregate LTV 0.69.

With pool-price triggers, liquidation sales depress prices, trigger further liquidations and raise collateral seizure per unit repaid. Holding liquidator incentives and pool settlement fixed, the pool-price trigger adds 432 and 110 USDC per common and balanced market at a 20 percent shock. It adds 674 and 477 at 30 percent. It alone produces exposure at 10 percent. That exposure is 51 USDC per common market, confined to four lever-up-first markets. Restoring the pool to $F$ after each liquidation round removes this exposure. It also lowers the 20 percent common-market exposure from 432 to 119 USDC.

With external triggers and pool settlement, exposure averages 272 USDC per common market and 226 per balanced market, but only 43 and 9 are written off. The remaining 229 and 217 are deficiencies on 2.9 and 4.6 accounts per market that can still be liquidated but that profit-seeking liquidators leave open. Write-offs alone would miss most uncovered claims.

\begin{table}[htbp]
\centering
\caption{\textbf{Conditional creditor exposure on the earlier experiment's books.} Deterministic counterfactuals with profit-seeking liquidators under the headline rules of Section~\ref{sec:rules} and the replenished pool-price rule. The first rows give pre-liquidation deficiency without liquidation. Values are USDC per market, with 12 markets per arm. Exposure/debt uses pooled arm debt. Table~\ref{tab:losses}, Panel B, gives the main experiment's books under the same rules.}
\label{tab:lossesorig}
\footnotesize
\setlength{\tabcolsep}{3.5pt}
\begin{tabular}{llrrrrrr}
\toprule
 & & \multicolumn{2}{c}{Creditor} & \multicolumn{2}{c}{of which} & \multicolumn{2}{c}{Exposure /} \\
 & & \multicolumn{2}{c}{exposure} & \multicolumn{2}{c}{remaining} & \multicolumn{2}{c}{pre-shock} \\
 & & \multicolumn{2}{c}{$U(F)$} & \multicolumn{2}{c}{deficiency} & \multicolumn{2}{c}{debt (\%)} \\
\cmidrule(lr){3-4}\cmidrule(lr){5-6}\cmidrule(lr){7-8}
Rule & Shock & C & H & C & H & C & H \\
\midrule
\textit{No liquidation ($U_0$)} & 10\% & 0 & 0 & -- & -- & -- & -- \\
 & 20\% & 0 & 0 & -- & -- & -- & -- \\
 & 30\% & 149 & 56 & -- & -- & -- & -- \\
\addlinespace
No feedback & 10\% & 0 & 0 & 0 & 0 & 0.0 & 0.0 \\
 & 20\% & 6 & 0 & 0 & 0 & 0.1 & 0.0 \\
 & 30\% & 471 & 371 & 0 & 0 & 5.6 & 4.2 \\
\addlinespace
External trigger, pool sale & 10\% & 0 & 0 & 0 & 0 & 0.0 & 0.0 \\
 & 20\% & 0 & 0 & 0 & 0 & 0.0 & 0.0 \\
 & 30\% & 272 & 226 & 229 & 217 & 3.2 & 2.6 \\
\addlinespace
Pool trigger, pool sale & 10\% & 51 & 0 & 0 & 0 & 0.6 & 0.0 \\
 & 20\% & 432 & 110 & 0 & 0 & 5.1 & 1.3 \\
 & 30\% & 946 & 703 & 0 & 0 & 11.3 & 8.0 \\
\addlinespace
\quad pool replenished & 10\% & 0 & 0 & 0 & 0 & 0.0 & 0.0 \\
 & 20\% & 119 & 3 & 0 & 0 & 1.4 & 0.0 \\
 & 30\% & 660 & 486 & 0 & 0 & 7.9 & 5.6 \\
\bottomrule
\end{tabular}

\end{table}

\paragraph{Common and balanced books.} Comparisons between common and balanced books are descriptive. On the earlier experiment's books, common-market mean exposure at a 30 percent shock exceeds the balanced mean under every headline rule. The paired sign-flip calculation gives $p=0.61$ for no feedback (471 against 371 USDC; 12 nonzero pairs). Of 44 common-market borrowers above the 0.648 write-off threshold, 34 are in the four lever-up-first markets. About 93 of the 100 USDC no-feedback gap comes from positions above 70 percent of reference value.

On the main experiment's books, the fixed-price gap is 41 USDC (0.0098 $K_0$). None of the 19 rules produces exposure at 10 percent, and three rules give lower common-market mean exposure at 30 percent. Exploratory worst-three common/balanced ratios range from 1.31 to 2.15 across the grid (1.93, 1.64 and 2.07 for the headline rules). On the earlier experiment's books they range from 2.3 to 2.4.

\paragraph{Paired comparisons.} The paired sign-flip test is exact under joint invariance of outcomes to independent within-block C/H label swaps. C and H roles were fixed, so this invariance is a model assumption. The two-sided test enumerates all 4,096 patterns. With $n > 0$ nonzero paired differences, its smallest attainable $p$ is $2/2^n$. With all differences zero, $p = 1$. For the three headline rules at 30 percent, all 12 pairs are nonzero. The $p$-values are 0.613 (no feedback), 0.643 (external trigger) and 0.579 (pool-price trigger). At 20 percent, $p$ is 0.250 for the no-feedback rule (3 nonzero pairs) and 0.133 for the pool-price trigger (10 pairs). At 10 percent, only the pool-price trigger has a nonzero pair, and $p$ is 0.125 (4 pairs). Across all 75 tested exposure comparisons, the smallest unadjusted $p$ is 0.125. Holm adjustment over the 18 headline tests fixed in the stress specification gives adjusted $p = 1.0$ for every exposure cell. These tests cover three rules, three shocks, exposure and liquidation transfer.

\paragraph{Normalization and concentration.} Pooled over markets, no-feedback exposure at 30 percent is 5.6 percent of pre-shock debt in common markets and 4.2 percent in balanced markets. Averaged market by market, it is 3.5 and 4.1 percent, because the delever-first common markets carry little debt and almost no exposure. The common arm's exposure sits in the lever-up-first markets. There, no-feedback scenario write-offs average 1,284 USDC, or 8.7 percent of pre-shock debt. Under the pool-price trigger at a 10 percent shock, four lever-up-first common markets have positive exposure (at most 513 USDC, 3.1 percent of that market's debt). No balanced market does. Common markets average 14.3 liquidations against 1.0. The worst three of 12 common markets are blocks 3, 4 and 8 under all three headline rules, and all are lever-up first. With no feedback they carry 1,342 USDC, against 566 for the worst three balanced markets (8.8 against 5.6 percent of their debt). The figures are 721 against 319 with an external trigger and 2,808 against 1,158 with a pool-price trigger. Across all 19 rules, the ratio of these worst-three means at 30 percent lies between 1.6 and 2.5. Leave-one-block-out ratios range from 2.08 to 2.46 (no feedback), 1.96 to 2.41 (external trigger) and 2.02 to 2.54 (pool-price trigger). Removing the four lever-up-first blocks reverses the worst-three comparison (132 against 554 USDC with no feedback).

\subsection{Fixed-request replay of the earlier experiment}\label{app:replay}

\begin{table}[htbp]
\centering
\caption{\textbf{Spot-valued versus reference-valued borrowing cap in a fixed-request replay of the earlier experiment.} The same recorded requests are executed under each cap, and agents do not react. $Y$ is in units of $K_0$, and $V_{20}$ is defined in the text. Peak run-up is measured on each run's own price path. Creditor exposure \eqref{eq:exposure} is at a 30 percent shock under three rules of this appendix. Deterministic counterfactual; descriptive; 12 blocks.}
\label{tab:a6}
\footnotesize
\begin{tabular}{p{7.0cm}ll}
\toprule
 & Spot-valued cap & Reference-valued cap \\
 & (recorded; replay exact) & (same requests, replayed) \\
\midrule
$Y$: mean, C / H & 0.59 / 0.64 & 0.46 / 0.56 \\
$Y$: SD, C / H (ratio) & 1.19 / 0.39 (3.02) & 1.00 / 0.34 (2.94) \\
$V_{20}$: SD ratio C/H & 2.65 & 2.61 \\
Largest borrower LTV at 100, C / H & 0.753 / 0.723 & 0.700 / 0.700 \\
Peak pool price run-up, lever-up-first C markets (\%, range) & 6.6--10.0 & 5.3--9.0 \\
Valid lever-ups executing zero, C / H & 116 / 143 & 143 / 159 \\
Executed lever-up tokens, C / H & 522 / 557 & 462 / 516 \\
End debt, mean per market, C / H (USDC) & 8,406 / 8,733 & 7,866 / 8,379 \\
\addlinespace
\multicolumn{3}{l}{\textit{Creditor exposure at a 30\% shock, C / H (USDC per market; \% of pre-shock debt)}} \\
\quad No feedback, mean & 471 / 371 (5.6 / 4.2) & 254 / 256 (3.2 / 3.1) \\
\quad\quad worst 3 of 12 & 1,342 / 566 & 679 / 365 \\
\quad External-price trigger, pool sale, profitable, mean & 272 / 226 (3.2 / 2.6) & 93 / 137 (1.2 / 1.6) \\
\quad\quad worst 3 of 12 & 721 / 319 & 184 / 173 \\
\quad\quad write-off/deficiency, C and H & 43/229 and 9/217 & 40/53 and 22/116 \\
\quad Pool-price trigger, pool sale, unconstrained, mean & 946 / 703 (11.3 / 8.0) & 654 / 548 (8.3 / 6.5) \\
\quad\quad worst 3 of 12 & 2,808 / 1,158 & 1,904 / 899 \\
\bottomrule
\end{tabular}

\end{table}

\begin{figure}[htbp]
\centering
\includegraphics[width=0.75\textwidth]{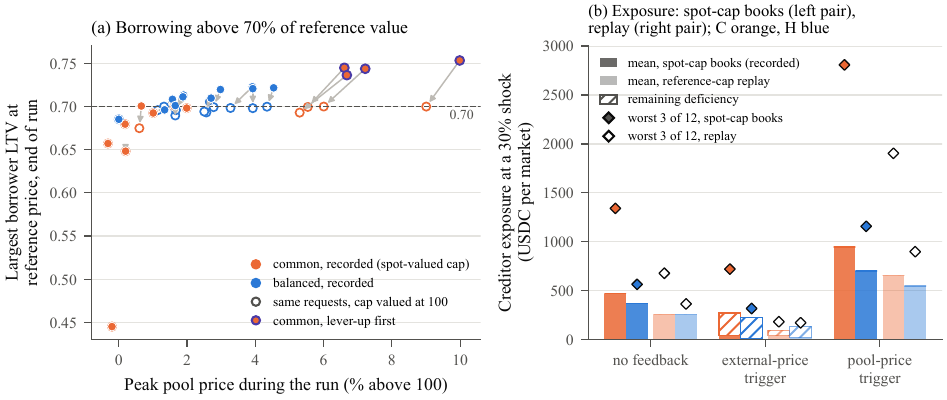}
\caption{\textbf{A reference-valued cap with requests held fixed (earlier experiment).} Each recorded request is re-executed in sequence, and agents do not adapt. (a) Peak end-of-round pool price against maximum terminal borrower reference LTV. Filled markers show the recorded spot-valued cap and hollow markers the reference-valued replay at its own peak price. Arrows join the same market, and violet outlines mark lever-up-first common markets. (b) Exposure at a 30 percent shock. Arm means are split into write-offs and remaining deficiency (hatched), with worst-three means as diamonds. The pool-price rule uses unconstrained liquidators, whose exposure equals that with profit-seeking liquidation. Table~\ref{tab:a6} gives the values.}
\label{fig:replay}
\end{figure}

The fixed-request replay of the earlier experiment compares cap valuations and recovery (Figure~\ref{fig:replay} and Table~\ref{tab:a6}). Under the recorded spot-valued cap, 32 borrowers in the four lever-up-first common markets exceeded 70 percent of reference collateral value while complying with the cap (Appendix~\ref{app:expb}). Under reference valuation none does, and pre-liquidation deficiency disappears. No-feedback exposure at 30 percent falls to 254 and 256 USDC per common and balanced market (3.2 and 3.1 percent of debt). All of it comes from the bonus on borrowers with LTVs between 0.648 and 0.70. Pool-price-trigger exposure remains 654 and 548 USDC. Reference valuation removes high-reference-LTV positions but leaves bonus-related write-offs and liquidation feedback.

The replay re-executes the earlier experiment's requests in recorded order and values the cap \eqref{eq:cap} at 100 rather than at the post-trade pool price. Agents do not observe the replay states or revise requests, so the comparison measures mechanical execution differences. Two independent implementations reproduce every execution, end-of-round price and end balance sheet under the spot-valued cap. One of them shares no code with the market simulator. They agree under the reference-valued cap to within $2 \times 10^{-11}$. A lever-up then executes the largest whole quantity that keeps the borrower's LTV, valued at 100, at or below 70 percent. Trade costs and pool dynamics follow the replayed path, and no liquidation occurs in any replayed market. $V_{20}$ is the debt in accounts whose reference LTV exceeds 0.64, divided by initial unused capacity. These accounts become liquidatable at a 20 percent shock. The stress results for the replayed books use the frozen stress engine with three rules at 20 and 30 percent. The rules are no feedback, the external trigger with profit-seeking liquidators and the pool-price trigger with unconstrained liquidators. As for the recorded books, the profit-seeking version of the pool-price rule gives exactly the same exposure on all 48 replayed books at both shocks.

\renewcommand{\floatpagefraction}{0.8}
\makeatletter\setlength{\@fptop}{0pt}\makeatother
\section{Historical comparison of the two market experiments}\label{app:history}

Both market experiments share the engine code and parameters of Section~\ref{sec:market}. They share the model families, mandates, slot rotation and model versions, with one explicit seed per decision. They also share the sample and within-block matching of Section~\ref{sec:randomization}, the allocation restrictions of Appendix~\ref{app:expb}, the prompt structure and feedback content of Section~\ref{sec:interfaces} and the outcome $Y$ (equation~\eqref{eq:y}). Table~\ref{tab:campaigns} documents what differs between them, and Table~\ref{tab:history} compares their recorded outcomes. The market experiments differ in interface, parser and feedback protocol. They also differ in books, seeds, schedules and allocation. I therefore compare them as historical descriptions.

Three exploratory counterparts apply the earlier experiment's definitions to the main experiment's records (Appendix~\ref{app:eexpl}). The first-action share of the common-market sum of squares is 68.1 percent, against 95.6 percent. The pooled within-group standard deviation in common markets is 0.29, against 0.23 across balanced markets. The earlier experiment had 0.27 against 0.39. Finally, 19 of the 38 common-market borrowers above the 0.648 write-off threshold at a 30 percent shock are in lever-up-first markets, compared with 34 of 44.

\begin{table}[htbp]
\centering
\caption{\textbf{Differences between the experiment designs.} Both market experiments use matched blocks. The text lists the features they share, and the two-panel table is available from the author on request.}
\label{tab:campaigns}
\footnotesize
\setlength{\tabcolsep}{4pt}
\begin{tabular}{@{}>{\raggedright\arraybackslash}p{0.19\linewidth}>{\raggedright\arraybackslash}p{\dimexpr0.405\linewidth-2\tabcolsep\relax}>{\raggedright\arraybackslash}p{\dimexpr0.405\linewidth-2\tabcolsep\relax}@{}}
\toprule
 & Earlier experiment (I0) & Main experiment (I1) \\
\midrule
Interface lines & Schema line lists the actions in the complete action order; ``If you hold, set qty 0.'' & Schema line lists no actions; ``Not trading is always a valid choice: choose hold, or set qty 0.'' \\
\addlinespace[2pt]
Zero quantity with lever-up or delever & Invalid output; nothing executed & Valid decision not to trade; the action label is kept \\
\addlinespace[2pt]
Feedback after no trade or an invalid output & ``you chose to hold''; an invalid output was ``treated as hold'' & ``you made no trade'' after any no-trade decision; after an invalid output, ``no trade was made'' \\
\addlinespace[2pt]
Books, schedules and seeds & Same distributions; drawn from this experiment's seed namespace & Same distributions; new draws from a separate namespace \\
\addlinespace[2pt]
Allocation draws & One draw per stage from a fixed-seed pseudo-random generator & One draw per stage from operating-system entropy; recorded, never redrawn \\
\addlinespace[2pt]
Invalid-output thresholds (11 per market, 54 per stage) & Stopping rule; it stopped the second stage after 8 of its 12 markets and was amended during the run into a reported diagnostic & Reported diagnostic from the start; no stopping or exclusion \\
\addlinespace[2pt]
Pre-specified test & Pitman--Morgan correlation of $Y^C + Y^H$ and $Y^C - Y^H$ over the 4,096 within-block swaps of the C and H labels; one-sided; null of joint exchangeability of the C and H outcomes & $T$, lever-up-first minus delever-first mean C$-$H, over 8,100 equally weighted plans of complete orders (balanced allocation fixed); two-sided; null that no common-market outcome changes under any reassignment \\
\addlinespace[2pt]
Stress analysis in the plan & None; a 19-rule grid was frozen for these books later, before any stress output & One fixed-price benchmark (30\% shock, no price feedback, 8\% bonus, completed liquidation) \\
\addlinespace[2pt]
Plan timing & Locked before the first call, after the frozen-state experiments; one in-run amendment (see above) & Locked before the first call; written after the results of the earlier experiment and the frozen-state interface test were known; one pre-live logging amendment; no live amendment, retry or replacement \\
\bottomrule
\end{tabular}

\end{table}

\begin{table}[htbp]
\centering
\caption{\textbf{Historical comparison of the market experiments.} $T$ uses the main experiment's definition. Its value for the earlier experiment is post hoc. Invalid and no-trade counts follow each market experiment's parser. Differences are computed before rounding.}
\label{tab:history}
\footnotesize
\begin{tabular}{lrr}
\toprule
 & Earlier & Main \\
 & experiment (I0) & experiment (I1) \\
\midrule
\multicolumn{3}{l}{\textit{Net debt growth $Y$ (units of $K_0$; 12 matched blocks)}} \\
$T$: lever-up-first minus delever-first mean of C $-$ H & 2.703 & 1.121 \\
Mean $Y$, common / balanced markets & 0.589 / 0.644 & 0.706 / 0.712 \\
Mean paired difference C $-$ H & $-$0.055 & $-$0.006 \\
SD of $Y$, common / balanced markets & 1.186 / 0.392 & 0.467 / 0.228 \\
SD ratio, common / balanced & 3.02 & 2.05 \\
\addlinespace
\multicolumn{3}{l}{\textit{Group means by first action of the common order (C / H / C $-$ H)}} \\
\quad lever-up first & 2.090 / 0.740 / 1.351 & 1.225 / 0.617 / 0.608 \\
\quad hold first & 0.238 / 0.401 / $-$0.163 & 0.497 / 0.609 / $-$0.112 \\
\quad delever first & $-$0.561 / 0.792 / $-$1.352 & 0.397 / 0.909 / $-$0.513 \\
\addlinespace
\multicolumn{3}{l}{\textit{Requests and execution (5,400 decisions per experiment)}} \\
Positive lever-up requests & 844 & 886 \\
\quad with any execution & 585 & 636 \\
\quad tokens requested / executed & 1,731 / 1,079 & 1,764 / 1,159 \\
Positive delever (collateral-sale) requests & 459 & 235 \\
\quad with any execution & 402 & 230 \\
\quad tokens requested / executed & 799 / 666 & 454 / 435 \\
Valid no-trade responses & 4,011 & 4,242 \\
Invalid outputs & 86 & 37 \\
Lever-up or delever with quantity 0 & 41 (invalid) & 51 (no trade) \\
\addlinespace
\multicolumn{3}{l}{\textit{Positive lever-up / delever requests by model family}} \\
\quad Llama & 418 / 246 & 732 / 235 \\
\quad Mistral & 307 / 213 & 154 / 0 \\
\quad Qwen & 119 / 0 & 0 / 0 \\
\addlinespace
\multicolumn{3}{l}{\textit{Losses (per unit of $K_0$; common / balanced markets)}} \\
Realized bad debt (no liquidation in any market) & 0 / 0 & 0 / 0 \\
Scenario write-off $B(70)$, fixed-price benchmark & 0.1104 / 0.0852 & 0.0705 / 0.0607 \\
Pre-liquidation deficiency $U_0(70)$ & 0.0351 / 0.0128 & 0.0128 / 0.0059 \\
\bottomrule
\end{tabular}

\end{table}

\let\FloatBarrier\relax

\section{Robustness and determinism checks}\label{app:audit}

These analyses use the recorded data and the existing simulator and make no model calls. They were chosen after the results of both market experiments were known and are exploratory.

\subsection{Determinism of the simulator}\label{app:falsify}

Deterministic baseline simulations confirm that the complete action order has zero effect in the absence of model responses.

\paragraph{Code trace.} The complete order is assigned to each agent when a market is set up. It is used only to render the prompt. In the rendered I1 prompt, exactly two lines differ between orders. They are the menu sentence and the quantity instruction. The request sent to the model contains the order only inside the prompt. The parser reads only the response text, and execution receives only the parsed action and quantity. Liquidation runs over borrowers in slot order. Arbitrage depends only on the pool state and fixed constants. Books depend on the block. The within-round schedule depends on the block and round, decision seeds on the block, slot and round, and news on the round. No execution-side quantity reads the order.

\paragraph{Design.} A copy of the market loop reproduces the recorded experiment. Fed the recorded responses, it reproduces all 24 markets of the main experiment exactly. This covers the 5,400 prompt hashes, every per-decision and per-round state, every final book and every $Y$. The language models are then replaced by three policies. Each policy first deletes the two order-carrying lines from the prompt and reads its inputs from the remaining text. The first policy always holds. Under the second, rule B, opportunistic borrowers lever up while their reference LTV is below 0.55. They buy the largest quantity of at most five tokens that the cap allows. Capital-preservation borrowers delever when the displayed LTV exceeds 45 percent. They sell $\min\{5,\,1+\lfloor(\text{LTV}-46)/5\rfloor\}$ tokens, at most the tokens held. Neutral borrowers lever up one token below 35 percent and delever one token above 60 percent. The third policy applies rule B and then cuts any active quantity above one by one token when a uniform draw from the decision's seed is below one half. This checks that seeds are consumed identically across orders. The thresholds are arbitrary. The rules serve only as order-blind decision makers. Each policy runs each of the 12 common-market books under all six complete orders, with every agent shown the same order. This gives 216 markets of 225 decisions. For rule B and its seeded variant, the state at every decision was also rendered under all six orders, and the policy's response was confirmed to be identical.

\paragraph{Results.} For each policy, every trajectory, final book and $Y$ is bit-identical across the six orders. The lever-up-first minus delever-first contrast is exactly zero (Table~\ref{tab:falsify}). No policy produces an invalid output, a blocked request or a liquidation. Under rule B, the books that the main experiment assigned to the lever-up-first and delever-first groups give mean $Y$ of 0.633 and 0.733. This difference of $-0.101$ reflects differences between the books, because the rule ignores the order. Its sign is opposite to that of the recorded gap.

\begin{table}[htbp]
\centering
\caption{\textbf{Order-blind deterministic policies under all six complete orders.} Mean $Y$ over the 12 main-experiment common books, each run under every order (216 markets). L first $-$ D first is the contrast of lever-up-first minus delever-first orders over all books. Largest spread is the maximum over books of the range of $Y$ across the six orders. E's order groups gives the same policy's mean $Y$ in the books assigned to lever-up-first common orders minus those assigned to delever-first orders in the main experiment. This column reflects differences between books only. Exploratory implementation check; no model calls.}
\label{tab:falsify}
\footnotesize
\setlength{\tabcolsep}{4pt}
\begin{tabular}{lrrrrrrrrr}
\toprule
 & \multicolumn{6}{c}{Mean $Y$ over the 12 common books under order} & L first & Largest & E's order \\
\cmidrule(lr){2-7}
Policy & LDH & LHD & DLH & DHL & HLD & HDL & $-$ D first & spread & groups \\
\midrule
Hold only & 0.0000 & 0.0000 & 0.0000 & 0.0000 & 0.0000 & 0.0000 & 0.0 & 0.0 & 0.000 \\
State rule B & 0.6792 & 0.6792 & 0.6792 & 0.6792 & 0.6792 & 0.6792 & 0.0 & 0.0 & $-$0.101 \\
B with seeded tie-break & 0.6352 & 0.6352 & 0.6352 & 0.6352 & 0.6352 & 0.6352 & 0.0 & 0.0 & $-$0.069 \\
\bottomrule
\end{tabular}

\end{table}

\subsection{Definitions and alternative path of the fixed-request decomposition}\label{app:decomp}

\paragraph{Environments.} Every environment executes each market's recorded requests. These are the recorded action and requested quantity, in the recorded order within each round. No-trade and invalid responses execute nothing. In A, a lever-up of $q$ tokens adds $100q$ of debt and $q$ tokens of collateral. A delever sells $\min(q,\lfloor c\rfloor)$ tokens at 100 and repays up to the outstanding debt. Any remainder goes to wallet cash. A has no cap, fee, price impact or liquidation. B adds a cap. A lever-up executes the largest whole $q'\le q$ with $d+100q'\le 0.70\cdot100\,(c+q')$. Post-trade LTV at a price of 100 never exceeds 0.70 in B, so the 80 percent liquidation rule cannot fire. X uses the recorded AMM (1,000 SMC and 100,000 USDC, 0.3 percent fee), per-round arbitrage that closes 25 percent of the log price gap and the recorded liquidation rule. It executes every lever-up in full at pool cost. C is X with the reference-valued cap. A lever-up executes the largest whole $q'$ with $d+\beta(q')\le 0.70\cdot100\,(c+q')$. D applies the recorded cap, equation~\eqref{eq:cap}. Wallet cash never constrains an action. Sales by borrowers without debt repay nothing but reduce collateral. $Y$ uses each block's recorded $D_b^0$ and $K_{0b}$.

\paragraph{Validation.} Environment D reproduces every recorded decision, round and final book (85,680 values, no difference) and every recorded $Y$ exactly. Environment C reproduces the reference-valued replay exactly. In every market and environment, net debt change equals borrowing less repayment, liquidation repayment and write-offs.

\paragraph{Results.} Table~\ref{tab:decompG} reports the common-market gap, the balanced-market gap and $T$ in each environment. $T$ is the common gap minus the balanced gap. The balanced-market gap is negative in every environment. The C$\to$D step for $T$ is 0.185. Of this, 0.141 comes from the common markets and 0.044 from a change in the balanced-baseline component. Net debt under D exceeds that under C in every common market, by 0.001 to 0.270 $K_0$. On the alternative path, AMM pricing without a cap lets every requested purchase execute at rising prices. This triggers 94 liquidation events and 98.74 USDC of write-offs across the 24 markets. Adding the reference-valued cap then raises the common gap by 0.127. The two paths share their endpoints but split the change differently.

\begin{table}[htbp]
\centering
\caption{\textbf{Fixed-request decomposition for common markets, balanced markets and $T$.} Lever-up-first minus delever-first gaps in mean $Y$ (units of $K_0$), four markets per group, in five environments. The lower rows give the changes along two paths. The main path runs A, B, C, D. The alternative path runs A, X, C, D. $T$ is the common gap minus the balanced gap. Computed from unrounded values. Exploratory fixed-request counterfactuals.}
\label{tab:decompG}
\footnotesize
\begin{tabular}{lrrr}
\toprule
 & Common-market gap & Balanced-market gap & $T$ \\
\midrule
\multicolumn{4}{l}{\textit{Level in each environment}} \\
A. Accounting at price 100, no cap & 1.549 & $-$0.356 & 1.905 \\
B. + reference-valued 70\% cap & 0.738 & $-$0.274 & 1.012 \\
X. AMM prices and arbitrage, no cap & 0.560 & $-$0.428 & 0.988 \\
C. + AMM prices and arbitrage & 0.688 & $-$0.248 & 0.936 \\
D. Pool-valued cap (recorded) & 0.828 & $-$0.293 & 1.121 \\
\addlinespace
\multicolumn{4}{l}{\textit{Changes along two paths}} \\
\quad A $\to$ B (main path) & $-$0.811 & 0.082 & $-$0.893 \\
\quad B $\to$ C (main path) & $-$0.051 & 0.025 & $-$0.076 \\
\quad C $\to$ D (main path) & 0.141 & $-$0.044 & 0.185 \\
\quad A $\to$ X (alternative path) & $-$0.989 & $-$0.071 & $-$0.918 \\
\quad X $\to$ C (alternative path) & 0.127 & 0.179 & $-$0.052 \\
\quad C $\to$ D (alternative path) & 0.141 & $-$0.044 & 0.185 \\
\bottomrule
\end{tabular}

\end{table}

\subsection{Debt level, dispersion and allocation}\label{app:alloc}

\paragraph{Metrics.} For each recorded end book, I compute four kinds of metrics over borrowers with positive collateral. Level metrics are aggregate reference LTV and the largest borrower LTV. Dispersion metrics are the Gini coefficient of borrower LTV (unweighted and collateral-weighted), the coefficient of variation of LTV and the HHI of debt shares (raw and normalized). Threshold metrics count the borrowers and the share of debt above reference LTVs of 0.648 and 0.70. They also include the debt above the 0.648 line, which equals the benchmark write-off $B(70)$. Exposure metrics are pre-liquidation deficiency $U_0(70)$ and exposure at a 30 percent shock under the three headline rules, computed with the stress engine.

\paragraph{Alternative allocations.} Holding each market's total debt and each borrower's collateral fixed, three rules reallocate debt. R1 allocates debt in proportion to collateral, which equalizes reference LTV. R2 allocates it in proportion to initial debt. R3 allocates it equally across borrowers with collateral. Each reallocated debt must satisfy the 70 percent cap at the reference price. Borrowers above the cap are set to it, and the excess is spread over the others in proportion to the rule's weights until no cap binds. R1 never needs capping. R2 and R3 are heavily constrained. In the main experiment's books, the cap moves 12,699 and 15,758 USDC and leaves 100 and 120 borrowers at the cap. In the recorded books, 39 main-experiment borrowers and 63 earlier-experiment borrowers end above the cap. Table~\ref{tab:allocrules} reports exposure under each allocation. R1 removes exposure under the fixed-price and external-trigger rules in all 24 main experiment's books. It leaves pool-trigger exposure only in two lever-up-first common markets (802 and 1,323 USDC). R2 and R3 remove pre-liquidation deficiency but keep 68 to 83 percent of the arm-mean benchmark write-off. Equal debt or initial-debt shares still put borrowers with little collateral above 0.648.

\begin{table}[htbp]
\centering
\caption{\textbf{Exposure under alternative debt allocations.} Mean USDC per market, with the number of markets with positive exposure in parentheses. There are 12 markets per arm. Shocks are to the reference price, with the recorded end-of-run pool reserves. $B(70)$ equals no-feedback exposure at 30 percent. The external and pool triggers use pool settlement, as in Table~\ref{tab:losses}. Borrowers at the cap and debt moved by the cap are totals over each market experiment's 24 books. For recorded books, the column counts borrowers above the cap, and no debt is moved. Total debt and each borrower's collateral are fixed in every allocation. Exploratory.}
\label{tab:allocrules}
\scriptsize
\setlength{\tabcolsep}{3.5pt}
\begin{tabular}{llrrrrrrr}
\toprule
 & & \multicolumn{2}{c}{Fixed price, $F = 70$} & External & \multicolumn{2}{c}{Pool trigger} & Borrowers & Debt moved \\
\cmidrule(lr){3-4}\cmidrule(lr){6-7}
Allocation & Arm & $B(70)$ & $U_0(70)$ & trigger, 30\% & 30\% & 20\% & at the cap & by the cap \\
\midrule
\multicolumn{9}{l}{\textit{A. Main experiment (I1)}} \\
Recorded & C & 323 (12) & 57 (9) & 215 (12) & 586 (12) & 134 (4) & 39 & 0 \\
 & H & 282 (12) & 27 (9) & 178 (12) & 507 (12) & 28 (7) & &  \\
R1 (equal LTV) & C & 0 (0) & 0 (0) & 0 (0) & 177 (2) & 0 (0) & 0 & 0 \\
 & H & 0 (0) & 0 (0) & 0 (0) & 0 (0) & 0 (0) & &  \\
R2 (initial debt shares) & C & 240 (11) & 0 (0) & 123 (11) & 510 (11) & 96 (4) & 100 & 12,699 \\
 & H & 234 (12) & 0 (0) & 135 (12) & 474 (12) & 23 (6) & &  \\
R3 (equal debt) & C & 240 (12) & 0 (0) & 144 (12) & 512 (12) & 86 (3) & 120 & 15,758 \\
 & H & 191 (12) & 0 (0) & 128 (12) & 370 (12) & 6 (3) & &  \\
\addlinespace
\multicolumn{9}{l}{\textit{B. Earlier experiment (I0)}} \\
Recorded & C & 471 (11) & 149 (5) & 272 (11) & 946 (11) & 432 (4) & 63 & 0 \\
 & H & 371 (12) & 56 (10) & 226 (12) & 703 (12) & 110 (8) & &  \\
R1 (equal LTV) & C & 77 (1) & 0 (0) & 11 (1) & 735 (4) & 151 (1) & 0 & 0 \\
 & H & 0 (0) & 0 (0) & 0 (0) & 17 (2) & 0 (0) & &  \\
R2 (initial debt shares) & C & 324 (11) & 0 (0) & 110 (11) & 874 (11) & 323 (5) & 137 & 25,554 \\
 & H & 294 (12) & 0 (0) & 165 (12) & 584 (12) & 49 (8) & &  \\
R3 (equal debt) & C & 302 (12) & 0 (0) & 113 (12) & 833 (12) & 312 (4) & 144 & 26,428 \\
 & H & 241 (12) & 0 (0) & 140 (12) & 490 (12) & 41 (5) & &  \\
\bottomrule
\end{tabular}

\end{table}

\paragraph{Level and distribution.} At each main-experiment book's initial aggregate LTV (0.36--0.42), the recorded distribution of debt produces no write-off or deficiency under the fixed-price and external-trigger rules. At the recorded aggregate LTV, the equal-LTV distribution produces none either. A decomposition that moves leverage first therefore attributes all exposure to leverage. One that moves the distribution first attributes all of it to the distribution. Leverage and distribution therefore interact completely. Under the pool-price trigger, leverage alone produces exposure in the two most levered lever-up-first common markets.

\paragraph{Reference-set positions.} Table~\ref{tab:allocmetrics} reports all 17 metrics for both market experiments. For each metric it gives the lever-up-first and delever-first common-market means and their difference. It also gives the paired version (C$-$H within block). As a placebo, it gives the same statistic for the balanced markets alone, which have no common order. ``Share'' is the fraction of the 8,100 first-action plans of Section~\ref{sec:estimand} whose statistic is at least as large in absolute value as the observed one, with the books held fixed. The smallest attainable value is $2/8{,}100\approx0.0002$. The metrics were chosen after the results were known. All 17 are reported, and several are functions of one another. Debt above the 0.648 line, $B(70)$ and no-feedback exposure coincide. The shares show where the observed assignment sits in the reference set.

\paragraph{Reading.} In the main experiment, the level and threshold metrics separate the two groups, with the same sign in both stages. The dispersion metrics are lower in lever-up-first markets on average, change sign between stages and have overlapping block ranges. In the earlier experiment, the level and threshold metrics separate the groups more sharply. The dispersion contrast there is large and reversed. Delever-first common markets ended with a few borrowers still levered and many at zero debt. Their LTVs are very unequal, while their aggregate leverage and exposure are low.

\begin{table}[p]
\centering
\caption{\textbf{Level, threshold, dispersion and exposure metrics of the recorded end books by first action.} All 17 metrics, for both market experiments. Means are per market over four common markets per group. Difference is L first minus D first. Paired is the within-block C$-$H difference. Balanced applies the same statistic to balanced markets grouped by their block's common order (placebo). Share is the fraction of the 8,100 first-action plans with an absolute statistic at least as large, books held fixed (minimum $2/8{,}100$). Shares are positions in the reference set. Exposure is at a 30 percent shock to the reference price, in USDC per market. Exploratory.}
\label{tab:allocmetrics}
\scriptsize
\setlength{\tabcolsep}{3.5pt}
\begin{tabular}{lrrrrrrr}
\toprule
 & \multicolumn{4}{c}{Common markets} & \multicolumn{2}{c}{Paired, C $-$ H} & Balanced \\
\cmidrule(lr){2-5}\cmidrule(lr){6-7}\cmidrule(lr){8-8}
Metric (per market) & L first & D first & Difference & Share & Difference & Share & Share \\
\midrule
\multicolumn{8}{l}{\textit{A. Main experiment (I1); L first in blocks 2, 5, 6, 11; D first in blocks 0, 1, 7, 9}} \\
Aggregate reference LTV & 0.559 & 0.470 & 0.089 & 0.0084 & 0.117 & 0.0005 & 0.2104 \\
Largest borrower LTV & 0.725 & 0.695 & 0.030 & 0.0035 & 0.041 & 0.0047 & 0.1751 \\
Gini of borrower LTV & 0.171 & 0.197 & $-$0.025 & 0.3640 & $-$0.040 & 0.1568 & 0.4963 \\
Gini of LTV, collateral-weighted & 0.148 & 0.178 & $-$0.029 & 0.1973 & $-$0.038 & 0.0215 & 0.5267 \\
Coefficient of variation of LTV & 0.319 & 0.373 & $-$0.055 & 0.2911 & $-$0.088 & 0.1094 & 0.4375 \\
HHI of debt shares & 0.098 & 0.104 & $-$0.007 & 0.5210 & $-$0.015 & 0.0640 & 0.2481 \\
HHI of debt shares, normalized & 0.033 & 0.035 & $-$0.002 & 0.8457 & $-$0.012 & 0.1881 & 0.2481 \\
Borrowers above 0.648 & 4.75 & 2.25 & 2.50 & 0.0531 & 3.50 & 0.0044 & 0.1889 \\
Debt share above 0.648 & 0.546 & 0.357 & 0.190 & 0.1481 & 0.244 & 0.0160 & 0.4980 \\
Borrowers above 0.70 & 4.00 & 0.50 & 3.50 & 0.0104 & 4.50 & 0.0030 & 0.2467 \\
Debt share above 0.70 & 0.469 & 0.078 & 0.391 & 0.0079 & 0.484 & 0.0020 & 0.3844 \\
Debt above the 0.648 line (USDC) & 630 & 151 & 479 & 0.0141 & 607 & 0.0010 & 0.1190 \\
Benchmark write-off $B(70)$ (USDC) & 630 & 151 & 479 & 0.0141 & 607 & 0.0010 & 0.1190 \\
Pre-liquidation deficiency $U_0(70)$ (USDC) & 158 & 3 & 155 & 0.0027 & 173 & 0.0022 & 0.2936 \\
Exposure, no feedback, 30\% (USDC) & 630 & 151 & 479 & 0.0141 & 607 & 0.0010 & 0.1190 \\
Exposure, external trigger, 30\% (USDC) & 350 & 132 & 219 & 0.0094 & 269 & 0.0017 & 0.2857 \\
Exposure, pool trigger, 30\% (USDC) & 1,210 & 229 & 981 & 0.0148 & 1,198 & 0.0002 & 0.1630 \\
\addlinespace
\multicolumn{8}{l}{\textit{B. Earlier experiment (I0), same metrics}} \\
Aggregate reference LTV & 0.642 & 0.341 & 0.300 & 0.0002 & 0.296 & 0.0002 & 0.9205 \\
Largest borrower LTV & 0.745 & 0.608 & 0.137 & 0.0002 & 0.139 & 0.0002 & 0.8133 \\
Gini of borrower LTV & 0.140 & 0.501 & $-$0.361 & 0.0002 & $-$0.331 & 0.0002 & 0.1262 \\
Gini of LTV, collateral-weighted & 0.107 & 0.317 & $-$0.211 & 0.0002 & $-$0.188 & 0.0007 & 0.2998 \\
Coefficient of variation of LTV & 0.271 & 0.965 & $-$0.694 & 0.0002 & $-$0.639 & 0.0002 & 0.1193 \\
HHI of debt shares & 0.093 & 0.189 & $-$0.096 & 0.0069 & $-$0.096 & 0.0086 & 0.9810 \\
HHI of debt shares, normalized & 0.028 & 0.119 & $-$0.091 & 0.0079 & $-$0.091 & 0.0089 & 0.9854 \\
Borrowers above 0.648 & 8.50 & 0.75 & 7.75 & 0.0015 & 8.75 & 0.0005 & 0.3385 \\
Debt share above 0.648 & 0.800 & 0.223 & 0.577 & 0.0040 & 0.727 & 0.0007 & 0.1677 \\
Borrowers above 0.70 & 8.00 & 0.00 & 8.00 & 0.0044 & 9.00 & 0.0017 & 0.4874 \\
Debt share above 0.70 & 0.773 & 0.000 & 0.773 & 0.0044 & 0.898 & 0.0020 & 0.4738 \\
Debt above the 0.648 line (USDC) & 1,284 & 14 & 1,271 & 0.0002 & 1,354 & 0.0010 & 0.5326 \\
Benchmark write-off $B(70)$ (USDC) & 1,284 & 14 & 1,271 & 0.0002 & 1,354 & 0.0010 & 0.5326 \\
Pre-liquidation deficiency $U_0(70)$ (USDC) & 448 & 0 & 448 & 0.0044 & 446 & 0.0037 & 0.9573 \\
Exposure, no feedback, 30\% (USDC) & 1,284 & 14 & 1,271 & 0.0002 & 1,354 & 0.0010 & 0.5326 \\
Exposure, external trigger, 30\% (USDC) & 687 & 14 & 673 & 0.0002 & 688 & 0.0007 & 0.8202 \\
Exposure, pool trigger, 30\% (USDC) & 2,656 & 21 & 2,634 & 0.0002 & 2,786 & 0.0010 & 0.6057 \\
\bottomrule
\end{tabular}

\end{table}

\subsection{Purchase cost and cross-borrower feasibility}\label{app:headroom}

\paragraph{The buyer's own room.} Let $r=1-\phi$, $t=q/x$ and $z=c_i/x$. The purchase cost and post-trade price of Section~\ref{sec:market} give $p'/p=(r+\phi t)/[r(1-t)^2]$. Substituting into equation~\eqref{eq:headroom} gives
\[
\Delta H_i=\frac{p\,q}{r(1-t)^2}\Big\{0.70\,z\,[(1+r)-rt]+0.70\,r-1+(1+0.70\,\phi)\,t\Big\}.
\]
The term in braces increases with $z$. The buyer's room therefore rises only if
\[
\frac{c_i}{x}\;>\;z^*(t)=\frac{1-0.70\,r-(1+0.70\,\phi)\,t}{0.70\,[(1+r)-rt]}.
\]
For a small purchase, $z^*(0)=(1-0.70r)/[0.70(1+r)]\approx0.216$. For purchases of one to five tokens in the initial pool, $z^*$ stays above 0.21. Pre-existing debt cancels from $\Delta H_i$. It affects only whether the purchase can execute. Another borrower $j$ whose position is unchanged gains $0.70\,c_j\,(p'-p)>0$.

\paragraph{A later borrower's feasible purchases.} Fix a later borrower with collateral $c$ and debt $d$ and a candidate purchase of $q$ tokens. Let $C=c+q$ and $w=x-q$. Under the pool-valued cap, the largest pre-existing debt that still allows this purchase is
\[
M(x,y)=0.70\,C\,p'(q)-\beta(q)=\frac{y\,N}{r\,w^2},\qquad N=(0.70\,rC-q)\,w+0.70\,Cq.
\]
The purchase is feasible exactly when $d\le M$. If it is feasible with $d\ge0$, then $N\ge0$ and
\[
\frac{\partial M}{\partial y}=\frac{M}{y}\ge0,\qquad \frac{\partial M}{\partial x}=-\frac{y\,(N+0.70\,Cq)}{r\,w^3}<0.
\]
An upstream purchase lowers $x$ and raises $y$. Along this change $N$ stays nonnegative, so $M$ rises over the whole change and not only locally. Suppose the later borrower's position is unchanged, reserves stay above $q$ and no sale, arbitrage or liquidation intervenes. Then every purchase that was feasible stays feasible with more slack. The engine executes the largest feasible quantity up to the request, so the executed quantity cannot fall. Under a reference-valued cap, the largest admissible debt is $M_{\rm ref}=0.70\,C\cdot100-\beta(q)$. An upstream purchase raises $\beta(q)$ and lowers $M_{\rm ref}$, so the later borrower's feasible set can only shrink.

\paragraph{Record check.} I apply equation~\eqref{eq:headroom} to the recorded states. All 636 executed purchases in the main experiment raised the pool price and lowered the buyer's own room, by 26 to 148 USDC. Buyers' collateral was 0.1 to 3.0 percent of the pool's tokens. Summed over all 15 borrowers, room fell after 618 purchases and rose after 18.

\section{Planned exploratory analyses}\label{app:planned}

This appendix reports three analyses planned after all results of both market experiments and the frozen-state test were known. The plan was hash-locked locally before any analysis script was written. The analyses use completed records, make no model calls and are exploratory.

\subsection{Order dependence of exposure and debt-adjustment endpoints}\label{app:plannedA}

For a market-level endpoint $Z$, the statistic has the pre-specified form. It is the mean of $Z_C-Z_H$ over lever-up-first blocks minus its mean over delever-first blocks. Its reference distribution uses the main experiment's 8,100 equally likely plans (Section~\ref{sec:inference}). The null is that every common-market value of $Z$ is invariant to every complete-order reassignment. Six endpoints were fixed in advance, and all are reported (Table~\ref{tab:plannedA}).

The sharp null is rejected for every endpoint. For benchmark write-offs and executed borrowing, the observed plan and its mirror are the two most extreme plans. Repayment gives the weakest evidence ($p=388/8{,}100$). Its common-market gap ($-0.351$) is similar in size to the borrowing gap (0.477). Most of the difference in evidence comes from the balanced baselines. They add 0.319 to $T_Z$ for borrowing and 0.026 for repayment. Write-offs and deficiency are scenario values on end-of-run books.

\begin{table}[htbp]
\centering
\caption{\textbf{Order dependence of exposure and debt-adjustment endpoints.} Main experiment. Exploratory sharp-null tests on the 8,100 plans of the pre-specified design, with the plan hash-locked locally. L first and D first give common-market means by the first action of the order. $T_Z$ applies the form of the pre-specified statistic to endpoint $Z$. Ref.\ SD is the SD of $T_Z$ over the reference set. The $p$-value is the share of plans with $|T_Z|$ at least the observed value. Holm adjusts across the six endpoints. Write-off and deficiency use the fixed-price benchmark at $F=70$ with an 8 percent bonus (Section~\ref{sec:ebenchmark}).}
\label{tab:plannedA}
\footnotesize
\setlength{\tabcolsep}{4pt}
\begin{tabular}{lrrrrrrr}
\toprule
 & \multicolumn{2}{c}{Common markets} & L$-$D & & Ref. & Two-sided & \\
\cmidrule(lr){2-3}
Endpoint & L first & D first & gap & $T_Z$ & SD & $p$ & Holm \\
\midrule
Benchmark write-off$/K_0$ & 0.139 & 0.033 & 0.106 & 0.135 & 0.052 & 2/8,100 & 0.0015 \\
Pre-liquidation deficiency$/K_0$ & 0.035 & 0.001 & 0.035 & 0.039 & 0.016 & 12/8,100 & 0.0044 \\
Executed borrowing$/K_0$ & 1.415 & 0.938 & 0.477 & 0.796 & 0.276 & 2/8,100 & 0.0015 \\
Repayment$/K_0$ & 0.190 & 0.541 & $-$0.351 & $-$0.325 & 0.172 & 388/8,100 & 0.0479 \\
Aggregate reference LTV & 0.559 & 0.470 & 0.089 & 0.117 & 0.041 & 4/8,100 & 0.0020 \\
Borrowers above 0.648 LTV & 4.8 & 2.2 & 2.5 & 3.5 & 1.4 & 36/8,100 & 0.0089 \\
\bottomrule
\end{tabular}

\end{table}

\subsection{State dependence of the order effect in frozen states}\label{app:plannedB}

The frozen-state interface test has 12 main states, four at each pool-valued LTV of 0.30, 0.50 and 0.66. Each state shows a pool price of 95 or 105 and no reference price. Price and cash are balanced within each LTV level. For each state and interface, I compute four order effects with equal cell weights (Table~\ref{tab:plannedB}). I then compare the mean effects across the three LTV levels. The plan also specified a permutation test of the LTV labels across the 12 states. I do not report it, because the labels were fixed by design and not randomly assigned.

Under I1 a request gap remained in the high-LTV states. The lever-up-first effect on the share of borrowing requests was 22, 11 and 13 points at the three LTV levels. The effect on executable tokens fell from 0.60 to 0.13, because the cap binds in the 0.66 states. There only a third of the requested-token contrast was executable. The delever-first effect on collateral-sale requests was 1, 15 and 17 points. For Llama it was 4, 29 and 50 points. Under I0 the effect on executable tokens also fell and the effect on sale requests also rose.

\begin{table}[htbp]
\centering
\caption{\textbf{Order effects by pool-valued LTV in frozen states.} Frozen-state interface test, exploratory. Each entry averages the state-level effect over the four states at that LTV. Lever-up outcomes compare lever-up first (LDH, LHD) with lever-up last (DHL, HDL). The sale outcome compares delever first (DLH, DHL) with delever last (LHD, HLD). Executable tokens are the tokens the cap would allow in that state. Difference is the mean at 0.66 minus the mean at 0.30.}
\label{tab:plannedB}
\footnotesize
\setlength{\tabcolsep}{4pt}
\begin{tabular}{lrrrr}
\toprule
 & \multicolumn{3}{c}{Order effect at pool-valued LTV} & \\
\cmidrule(lr){2-4}
Outcome & 0.30 & 0.50 & 0.66 & Difference \\
\midrule
\multicolumn{5}{@{}l}{\emph{Revised interface I1}} \\
Positive lever-up request & 0.222 & 0.111 & 0.125 & $-$0.097 \\
Requested lever-up tokens & 0.597 & 0.389 & 0.375 & $-$0.222 \\
Executable lever-up tokens & 0.597 & 0.389 & 0.125 & $-$0.472 \\
Positive collateral-sale request & 0.014 & 0.153 & 0.167 & 0.153 \\
\multicolumn{5}{@{}l}{\emph{Original interface I0}} \\
Positive lever-up request & 0.250 & 0.306 & 0.319 & 0.069 \\
Requested lever-up tokens & 0.681 & 0.556 & 0.472 & $-$0.208 \\
Executable lever-up tokens & 0.681 & 0.556 & 0.319 & $-$0.361 \\
Positive collateral-sale request & 0.083 & 0.319 & 0.306 & 0.222 \\
\bottomrule
\end{tabular}

\end{table}

\subsection{Behavior after blocked trades}\label{app:plannedC}

Table~\ref{tab:plannedC} reports the next decision of each Llama and Mistral borrower by its own previous decision. Blocked borrowers are closer to the cap. The right panel therefore keeps next decisions with remaining room of at most 10 percent of the borrower's pool-valued capacity.

In this panel, Llama followed a fully executed lever-up with another lever-up request in 70 percent of next decisions and with a sale in 5 percent. After a blocked lever-up, the rates were 46 and 22 percent. Mistral requested another lever-up after half of its blocked requests and never sold collateral.

\begin{table}[htbp]
\centering
\caption{\textbf{Next decision after one's own previous decision.} Main experiment, descriptive. Shares of next decisions with a positive lever-up request or a positive collateral-sale request. Room is the remaining borrowing room at the pool price as a share of pool-valued capacity $0.70\,c\,p$. Qwen made no trade.}
\label{tab:plannedC}
\footnotesize
\setlength{\tabcolsep}{4pt}
\begin{tabular}{lrrrrrr}
\toprule
 & \multicolumn{3}{c}{All next decisions} & \multicolumn{3}{c}{Room $\le$ 10\% of capacity} \\
\cmidrule(lr){2-4}\cmidrule(lr){5-7}
Own previous decision & $n$ & Lever-up & Sale & $n$ & Lever-up & Sale \\
\midrule
\multicolumn{7}{@{}l}{\emph{Llama}} \\
Lever-up fully executed & 504 & 0.667 & 0.052 & 155 & 0.697 & 0.052 \\
Lever-up partially executed & 38 & 0.605 & 0.105 & 38 & 0.605 & 0.105 \\
Lever-up blocked & 153 & 0.464 & 0.229 & 147 & 0.463 & 0.225 \\
Collateral sale & 219 & 0.301 & 0.429 & 49 & 0.408 & 0.408 \\
No positive lever-up & 766 & 0.231 & 0.084 & 231 & 0.242 & 0.078 \\
\multicolumn{7}{@{}l}{\emph{Mistral}} \\
Lever-up fully executed & 63 & 0.905 & 0.000 & 25 & 0.960 & 0.000 \\
Lever-up partially executed & 8 & 0.125 & 0.000 & 8 & 0.125 & 0.000 \\
Lever-up blocked & 74 & 0.500 & 0.000 & 74 & 0.500 & 0.000 \\
No positive lever-up & 1,535 & 0.029 & 0.000 & 71 & 0.507 & 0.000 \\
\bottomrule
\end{tabular}

\end{table}

\section{Larger-model studies and data availability}\label{app:other}

\subsection{The larger-model frozen-state test}\label{app:larger}

I ran the frozen-state interface test of Appendix~\ref{app:expd} on three larger configurations, with the plan and code hash-locked locally before the first call. They are Qwen 2.5 14B (local), Llama 3.2 90B Vision (hosted by NVIDIA, text only, with the Llama 3.1 70B text model) and Gemini 4 Argon (Vertex AI, low thinking level). A locked amendment adopted this Llama model before any Llama grid call, because NVIDIA had retired Llama 3.3 70B. Each model received the same 432 prompts, and Table~\ref{tab:larger} reports each model separately.

\begin{table}[htbp]
\centering
\caption{\textbf{Larger-model frozen-state test.} $E_{I0}$ and $E_{I1}$ give the rate of positive lever-up requests with lever-up listed first minus the rate with it listed last, in percentage points. Brackets give 95 percent state-cluster bootstrap intervals (10,000 draws). $\Delta$ is $E_{I1}-E_{I0}$, and its $p$-value is the exact sign-flip value over the 12 states. Equal cell weights. Missing cells had no usable response after the allowed re-sends and are excluded, as the plan specified. Intervals of zero width mean that no cell had a positive request. Prompts contain only synthetic balance sheets and mandate text. Vertex AI's paid or trial terms do not use them for training, and NVIDIA's evaluation terms may. Re-issuing the calls of one state kept the decision category in every pair received.}
\label{tab:larger}
\footnotesize
\setlength{\tabcolsep}{3.5pt}
\begin{tabular}{@{}lcccccccc@{}}
\toprule
Model & Cells used & Missing & No trade (\%) & $E_{I0}$ & $E_{I1}$ & $\Delta$ & $p(\Delta)$ \\
 & of 432 & I0 / I1 & I0 / I1 & & & & \\
\midrule
Qwen 2.5 14B & 432 & 0 / 0 & 86.1 / 100.0 & 15.3 [6.9, 23.6] & 0.0 [0.0, 0.0] & $-15.3$ [$-23.6$, $-6.9$] & 0.016 \\
Llama 3.2 90B Vision & 399 & 14 / 19 & 100.0 / 100.0 & 0.0 [0.0, 0.0] & 0.0 [0.0, 0.0] & 0.0 [0.0, 0.0] & 1.000 \\
Gemini 4 Argon & 427 & 2 / 3 & 100.0 / 100.0 & 0.0 [0.0, 0.0] & 0.0 [0.0, 0.0] & 0.0 [0.0, 0.0] & 1.000 \\
\bottomrule
\end{tabular}

\end{table}

In these static, newsless states, trading activity varies across the models. Qwen 14B made positive lever-up requests in 5.1 percent of I0 cells and positive sale requests in 8.8 percent. Its first-listed effect was 15.3 points on lever-up requests and 18.1 points on sale requests. Under I1 it chose no trade in all 216 cells. Llama 90B and Gemini 4 Argon chose no trade in every cell that returned a response, under both interfaces. A shared order reaches credit through trades, so the next subsection turns to a market in which these models trade.

\subsection{Markets with larger models and price-relevant news}\label{app:news}

A second study reran the market of the main experiment with price-relevant news. News arrives in rounds 3, 6, 9 and 12 with signs $+,-,+,-$. Each item moves the reference price by a log step of $m=0.05$, chosen by a locked rule in a 648-call calibration. The \emph{large} campaign places Llama 90B, Qwen 14B and Gemini 4 Argon in the three model positions, and the \emph{small} campaign uses the three models of the main experiment. Each campaign has 24 markets and 5,400 decisions and reuses the books, seeds and realized allocation of the main experiment. The plan was hash-locked locally before the first market call. Table~\ref{tab:news} gives all pre-specified tests, and the complete results are available from the author on request.

\begin{table}[htbp]
\centering
\caption{\textbf{News-market study.} Panel A applies the statistic and the 8,100 plans of the main experiment within each campaign. Holm adjusts across the four secondary endpoints. Net debt growth, borrowing, repayment and write-offs are in units of $K_0$. Write-offs use the fixed-price benchmark. Aggregate LTV is in LTV units. Panel B is pre-specified descriptive and gives the share of decisions with a positive request.}
\label{tab:news}
\footnotesize
\setlength{\tabcolsep}{4pt}
\begin{tabular}{@{}lrrrrrr@{}}
\toprule
\multicolumn{7}{@{}l}{\emph{Panel A. Pre-specified tests}} \\
 & \multicolumn{3}{c}{Large campaign} & \multicolumn{3}{c}{Small campaign} \\
\cmidrule(lr){2-4}\cmidrule(l){5-7}
Endpoint & $T$ & $p$ & Holm & $T$ & $p$ & Holm \\
\midrule
Net debt growth (primary) & 0.330 & 162/8,100 & & 1.328 & 2/8,100 & \\
Executed borrowing & 0.221 & 0.003 & 0.012 & 0.835 & 0.009 & 0.017 \\
Repayment & $-0.109$ & 0.166 & 0.237 & $-0.494$ & 0.025 & 0.025 \\
Benchmark write-off & 0.009 & 0.119 & 0.237 & 0.169 & 0.004 & 0.011 \\
Aggregate reference LTV & 0.055 & 0.022 & 0.064 & 0.171 & 0.0002 & 0.001 \\
\midrule
\multicolumn{7}{@{}l}{\emph{Panel B. Share of decisions with a positive request}} \\
 & \multicolumn{3}{c}{Large campaign} & \multicolumn{3}{c}{Small campaign} \\
\cmidrule(lr){2-4}\cmidrule(l){5-7}
 & Llama 90B & Qwen 14B & Gemini & Llama 8B & Qwen 7B & Mistral 7B \\
\midrule
News rounds & 0.79 & 0.85 & 0.49 & 0.88 & 0.09 & 0.66 \\
Other rounds & 0.002 & 0 & 0 & 0.60 & 0 & 0.12 \\
\bottomrule
\end{tabular}

\end{table}

Both campaigns reject complete-order invariance. In a descriptive comparison, the large-campaign estimate is about a quarter of the small-campaign estimate on the same schedule. Paired C$-$H means are $+0.155$ in lever-up-first blocks and $-0.175$ in delever-first blocks of the large campaign (4 of 4 positive; 3 of 4 negative and one near zero). They are $+0.683$ and $-0.646$ in the small campaign. Among the secondary endpoints, only borrowing passes the Holm adjustment at 0.05 in the large campaign. All four pass in the small campaign.

Exploratory accounting splits locate the large-campaign estimate. By model, Llama 90B contributes $+0.262$, Qwen 14B $+0.066$ and Gemini $+0.002$. By round, news rounds contribute $+0.324$ and other rounds $+0.006$. The larger models made 2 positive requests in 3,960 decisions outside news rounds. In the small campaign the split by round is $+0.650$ and $+0.678$.

Neither campaign had technical missing decisions or liquidations. The large campaign had no invalid output, and the small campaign had 64, which executed as no trade.

\subsection{Data and code availability}\label{app:materials}

The locally hash-locked plans and their hash records are available from the author on request. So are the call ledgers, raw responses and the decision, round and final-book records. The simulator, interfaces, parsers and analysis scripts are also available, along with the stored outputs behind the reported tables and figures and the complete results of the studies in this appendix. The scripts regenerate the analyses from the stored records.

\end{document}